\documentclass[twocolumn]{aastex63}

\usepackage{amsmath}
\usepackage{amssymb}
\usepackage{graphicx}
\usepackage{xcolor}
\usepackage{multirow}

\newcommand{\Om}{\Omega_\mathrm{m}}
\newcommand{\Ob}{\Omega_\mathrm{b}}
\newcommand{\s}[1]{\sigma_{#1}}
\newcommand{\ee}[1]{\times10^{#1}}
\newcommand{\Msun}{\mathrm{M}_\odot}
\newcommand{\unit}[1]{\,\mathrm{#1}}
\newcommand{\hMpc}{h^{-1}\mathrm{Mpc}}

\shorttitle{Cosmology with HSC first-year data using DL}
\shortauthors{Lu et~al.}

\begin{document}

\title{Cosmological constraints from HSC survey first-year data using deep learning}

\correspondingauthor{Tianhuan Lu}
\email{tl2854@columbia.edu}

\author[0000-0003-1040-2639]{Tianhuan Lu}
\affiliation{Department of Astronomy, Columbia University, New York, NY 10027, USA}

\author[0000-0003-3633-5403]{Zolt\'an Haiman}
\affiliation{Department of Astronomy, Columbia University, New York, NY 10027, USA}
\affiliation{Department of Physics, Columbia University, New York, NY 10027, USA}

\author[0000-0003-2880-5102]{Xiangchong Li}
\affiliation{Department of Physics, McWilliams Center for Cosmology, Carnegie Mellon University, Pittsburgh, PA 15213, USA}

\begin{abstract}
We present cosmological constraints from the Subaru Hyper Suprime-Cam (HSC) first-year weak lensing shear catalogue using convolutional neural networks (CNNs) and conventional summary statistics. We crop 19 $3\times3\unit{{deg}^2}$ sub-fields from the first-year area, divide the galaxies with redshift $0.3\le z\le1.5$ into four equally-spaced redshift bins, and perform tomographic analyses. We develop a pipeline to generate simulated convergence maps from cosmological $N$-body simulations, where we account for effects such as intrinsic alignments (IAs), baryons, photometric redshift errors, and point spread function errors, to match characteristics of the real catalogue. We train CNNs that can predict the underlying parameters from the simulated maps, and we use them to construct likelihood functions for Bayesian analyses. In the $\Lambda$ cold dark matter model with two free cosmological parameters $\Om$ and $\s8$, we find $\Om=0.278_{-0.035}^{+0.037}$, $S_8\equiv(\Om/0.3)^{0.5}\s8=0.793_{-0.018}^{+0.017}$, and the IA amplitude $A_\mathrm{IA}=0.20_{-0.58}^{+0.55}$. In a model with four additional free baryonic parameters, we find $\Om=0.268_{-0.036}^{+0.040}$, $S_8=0.819_{-0.024}^{+0.034}$, and $A_\mathrm{IA}=-0.16_{-0.58}^{+0.59}$, with the baryonic parameters not being well-constrained. We also find that statistical uncertainties of the parameters by the CNNs are smaller than those from the power spectrum (5--24 percent smaller for $S_8$ and a factor of 2.5--3.0 smaller for $\Om$), showing the effectiveness of CNNs for uncovering additional cosmological information from the HSC data. With baryons, the $S_8$ discrepancy between HSC first-year data and Planck 2018 is reduced from $\sim2.2\,\sigma$ to $0.3\text{--}0.5\,\sigma$.
\end{abstract}


\section{Introduction}
\label{sec:introduction}

According to general relativity, the light originating from a distant galaxy is bent by inhomogeneities in the foreground
matter distribution, which typically leads to a small distortion to the apparent shape of the galaxy -- an effect called weak lensing. With the measurements of millions of galaxy shapes, we can reconstruct weighted projections of the matter density in our Universe and use these to infer the underlying cosmological model. Weak lensing has been proposed to be a powerful tool to constrain some of the cosmological parameters, such as the mean matter density $\Om$ and the normalisation of matter fluctuation $\s8$  in a $\Lambda$ cold dark matter ($\Lambda$CDM) universe \citep[see, e.g.][for reviews]{bartelmann2001,refregier2003,kilbinger2015}, among many other probes such as the cosmic microwave background \citep[e.g.][]{hinshaw2013, planck2018} and baryon acoustic oscillations \citep[e.g.][]{anderson2014,alam2017}. The current weak lensing surveys are referred to as the ``Stage III'' surveys, such as the Dark Energy Survey \citep[DES;][]{abbott2016}, the Hyper Suprime-Cam (HSC) survey \citep{aihara2018a}, and the Kilo-Degree survey \citep[KiDS;][]{kuijken2015}. Due to their large survey area and depth, these surveys are currently able to yield constraints on the parameter $S_8\equiv(\Om/0.3)^{0.5}\s8$ with $\lesssim5\text{ percent}$ relative error \citep[see e.g.][]{hikage2019,hamana2020,heymans2021,abbott2022}.

In general, cosmological parameters are constrained by fitting the observation to a theoretical model through some summary statistics. The most widely used of these are two-point statistics such as the two-point correlation function (2PCF) and the power spectrum of the lensing shear, which are considered optimal in extracting Gaussian information from weak lensing signals \citep[see e.g.][]{fu2014,kohlinger2016,hikage2019,hamana2020,heymans2021,abbott2022}. One limitation of these two-point statistics is their inability to fully utilise non-Gaussian information on small, non-linear scales (below a few arcmin). Hence, many non-Gaussian summary statistics have been proposed and shown certain improvements over the 2PCF and power spectrum, e.g. three-point functions \citep{takada2003,vafaei2010}, bispectra \citep{takada2004,dodelson2005}, peak counts \citep{jain2000a,dietrich2010,kratochvil2010,kilbinger2015,liu2015,martinet2018}, and Minkowski functionals \citep{munshi2011,kratochvil2012,petri2013}.

An alternative approach to extracting information from weak lensing signals is to use machine learning, where an algorithm is designed to identify features from the signal, usually in the form of lensing shear maps or convergence maps, and uses them to estimate the parameters. Convolutional neural networks \citep[CNNs;][]{lecun1998} are considered one of the promising methods due to their success in image classification and object detection tasks \citep{krizhevsky2012,he2016,ren2015,redmon2016}.  In the context of weak lensing, multiple recent studies have investigated the use of CNNs (with different network architectures) and shown that they can outperform handcrafted summary statistics. Using suites of idealised noiseless simulated convergence maps, \citet{gupta2018} showed that the CNN achieved $\Om\text{--}\s8$ constraints that are five times tighter than those from the power spectrum. \citet{ribli2019} found $2.4–2.8$ times tighter constraints from the CNNs than the power spectrum on noisy maps in an LSST-like survey. In the first application to real observations, \citet{fluri2019} analysed  the KiDS-450 survey and derived a 30 percent improvement in constraining $S_8$ by using the CNNs compared to the power spectrum.  \citet{fluri2022} subsequently found a 16 percent improvement using KiDS-1000 data. 

Despite yielding tighter constraints, using CNNs to extract information from weak lensing signals comes with two challenges. First, we need to establish a pipeline to simulate the lensing shear maps or convergence maps. Usually, a large number of cosmological simulations that cover the whole parameter space need to be run, and mock lensing maps that match the target survey are generated from them. This is unlike the 2PCF and power spectrum, for which theoretical predictions can be calculated from the non-linear matter power spectrum calibrated by a small number of simulations \citep[e.g.][]{kilbinger2009,takahashi2012}. Second, since we have very little control over what features the CNNs are focusing on, we need to carefully account for the systematic effects on a pixel-by-pixel level so that the CNNs will not make predictions based on features that are wrongly presented in the simulated maps.

In this study, we perform weak lensing analyses applying CNNs to data from the HSC Subaru Strategic Program \citep[HSC SSP;][]{aihara2018b}.  $137\unit{{deg}^2}$ of galaxy shear measurements are currently publicly available from this survey. In our simulation pipeline, we produce mock shear catalogues and convergence maps based on a large $N$-body simulation suite, which then incorporates systematic effects including the intrinsic alignments, baryonic effects, photometric redshift estimation uncertainties, shear measurement biases, and point spread function (PSF) modelling errors.

This paper is organised as follows.
In Section~\ref{sec:catalog}, we describe how we prepare the HSC first-year catalogue, as well as the generation of mock catalogues from the simulations by \citet{takahashi2017} (T17 hereafter). 
In Section~\ref{sec:simulation}, we present our own simulation pipeline in detail, including the $N$-body simulations, ray-tracing, baryonic model, intrinsic alignments, and treatments of other systematics. 
In Section~\ref{sec:statistics}, we describe the calculation of conventional summary statistics---power spectrum, and peak counts---and the CNNs, which we argue can be regarded as a different type of summary statistic. 
We then describe the process of parameter inference using Bayesian analyses in Section~\ref{sec:parameter-inference}, and we present the cosmological constraints using various models in Section~\ref{sec:constraints}. 
In Section~\ref{sec:discussion}, we discuss the impact of CNN hyper-parameters, the choice of smoothing scale, the origin of non-Gaussian information, and we compare our constraints to Planck 2018 results. Finally, we summarise our main conclusions in Section~\ref{sec:conclusions}.

\section{catalogue preparation}
\label{sec:catalog}

\subsection{HSC first-year catalogues}
\label{sec:hsc-catalog}

\begin{figure*}[!t]
\centering
\includegraphics[width=17cm]{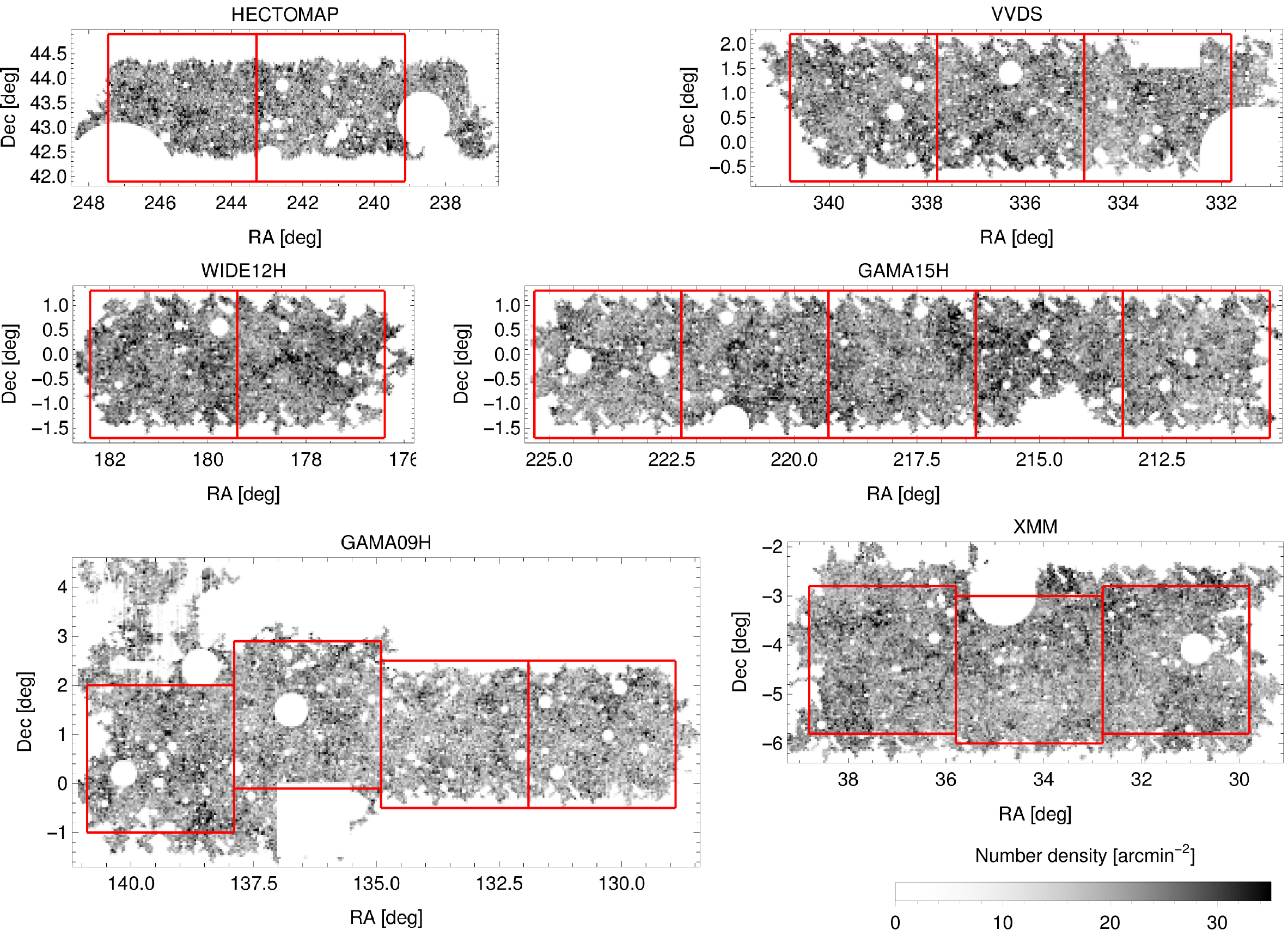}
\caption{The galaxy number density in the HSC first-year shear catalogue. The red squares shows the 19 sub-fields cropped from the catalogue and used in forward modelling with our ray-tracing simulations.}
\label{fig:fields}
\end{figure*}

\begin{figure}[!t]
\centering
\includegraphics[width=7.5cm]{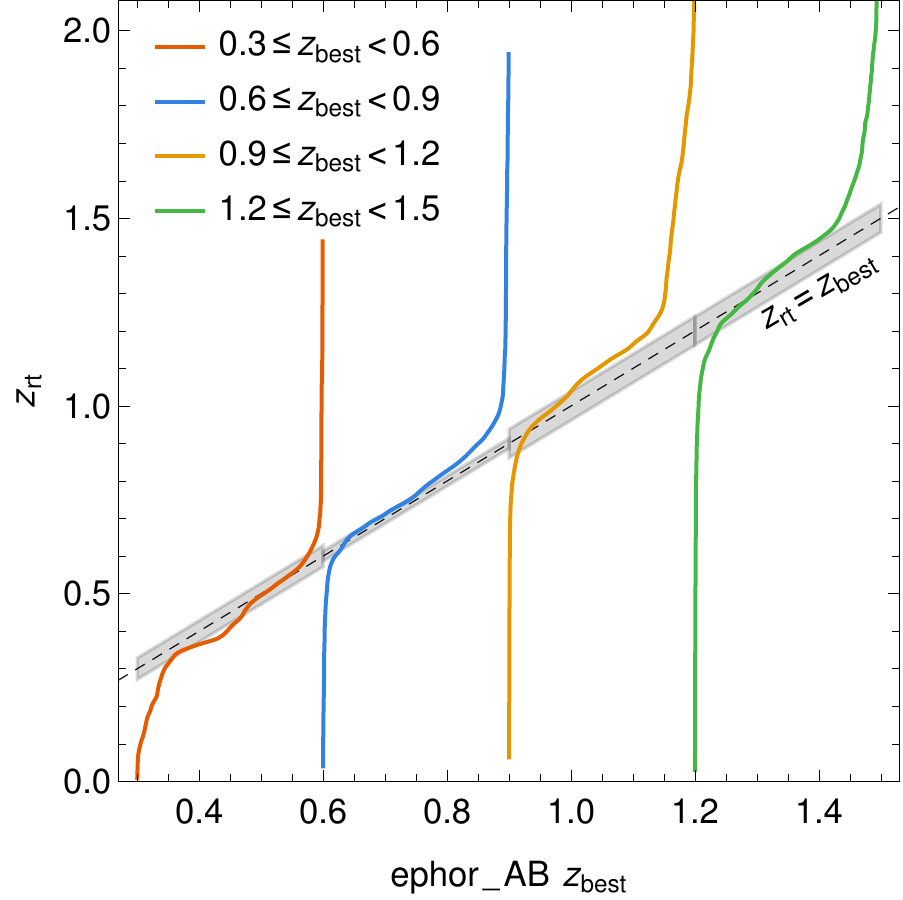}
\caption{The relation between \texttt{ephor\_AB} $z_\mathrm{best}$ and the assigned ray-tracing redshifts $z_\mathrm{rt}$ for each redshift bin. The light grey shade in each bin shows the $\pm1\sigma_{\Delta z}$ range around $z_\mathrm{rt}=z_\mathrm{best}$.}
\label{fig:z-rt}
\end{figure}

The cosmological inference in this work is based on the HSC first-year shear catalogue prepared by \citet{mandelbaum2018a}. Apart from masked regions due to bright stars, the catalogue covers $136.9\unit{{deg}^2}$ of sky area with an unweighted galaxy number density of $24.6\unit{{arcmin}^{-2}}$ \citep{mandelbaum2018a}, as is shown in Figure~\ref{fig:fields}. To facilitate our $N$-body simulation and ray-tracing pipeline, we crop 19 sub-fields from the catalogue, each of which can be projected onto a $3\times3\unit{{deg}^2}$ map using the flat sky projection \citep{coxeter1989, liu2015}:
\begin{eqnarray}
x&=&r^{-1}\cos\alpha\sin(\delta-\delta_0), \\
y&=&r^{-1}\left(\cos\alpha_0\sin\alpha-\sin\alpha_0\cos\alpha\cos(\delta-\delta_0)\right), \\
r&=&\sin\alpha_0\sin\alpha+\cos\alpha_0\cos\alpha\cos(\delta-\delta_0),
\end{eqnarray}
where $(\alpha, \delta)$ is the $(\mathrm{RA}, \mathrm{Dec})$ coordinate of the galaxy, $(\alpha_0, \delta_0)$ the center of the sub-field, and $(x, y)$ the projected position of the galaxy. Figure~\ref{fig:fields} shows the boundaries of these sub-fields in red squares.

To perform tomographic analyses, we gather the photometric redshifts (photo-$z$) of the galaxies from the HSC photo-$z$ catalogue \citep[see][]{tanaka2018}. Following \citet{hikage2019}, we put the galaxies between $z=0.3$ and $z=1.5$ into four equally-spaced redshift bins according to their \texttt{ephor\_AB} $z_\mathrm{best}$ from the catalogue. The galaxies outside this redshift range will not be used in this paper. Out of 11.9 million galaxies in the original catalogue, 29 percent are discarded by sub-field cropping (5 percent) and redshift binning (24 percent), leaving us with 8.5 million galaxies.

Additionally, our pipeline is dependent on the redshift distribution of the galaxies within each bin. Note that although the $z_\mathrm{best}$ estimations are considered \emph{best} for individual galaxies, the distribution of all $z_\mathrm{best}$ estimations is a biased estimate of the true redshift distribution, which can degrade the cosmological constraints \citep[see e.g.][]{abruzzo2019}. To address this issue, we adopt the COSMOS \citep{laigle2016} re-weighting method from \citet{hikage2019} to get reliable redshift distributions. The weighted COSMOS photo-$z$ catalogue\footnote{https://hsc-release.mtk.nao.ac.jp/doc/index.php/s17a-wide-cosmos/} includes 30-band photo-$z$ estimations, much more accurate than the HSC 5-band photo-$z$, but they are only available for a small number of galaxies compared to the full shear catalogue. So each of those galaxies is assigned a weight such that the colour magnitude distribution of all galaxies matches that in the first-year shear catalogue. For the $b^\text{th}$ redshift bin ($b=1,2,3,4$), we pick the representative galaxies that are both in the $b^\text{th}$ bin and in the COSMOS catalogue, and we consider their weighted 30-band photo-$z$ distribution to be a good estimation for the true redshift distribution $p_b(z)$.

With $z_\mathrm{best}$ and $p_b(z)$, we now assign \emph{ray-tracing redshifts} ($z_\mathrm{rt}$) to the galaxies with the following steps. First, we randomly sample $N_b$ candidate $z_\mathrm{rt}$ values from $p_b(z)$, where $N_b$ is number of galaxies in the bin. Then, the $i^\text{th}$ lowest $z_\mathrm{rt}$ is assigned to the galaxy with the $i^\text{th}$ lowest $z_\mathrm{best}$ for $i=1,2,\cdots,N_b$. Such assignments ensure that the distribution of $z_\mathrm{rt}$ reproduces the true redshift distribution of galaxies within the bin, and since $z_\mathrm{rt}$ has the same ordering as $z_\mathrm{best}$, the distance information is partially retained. Figure~\ref{fig:z-rt} shows the mappings from $z_\mathrm{best}$ to $z_\mathrm{rt}$. 
Compared to an alternative approach where the $z_\mathrm{rt}$ are assigned randomly without considering $z_\mathrm{best}$, we find the difference in the statistical properties of the lensing signal to be negligible for HSC first-year data according to our tests.

\subsection{T17 mock shear catalogues}
\label{sec:mock-catalogues}

In parallel with the cosmological analyses on the real HSC data, we perform analyses on mock catalogues based on the $N$-body simulations and full-sky ray-tracing by T17. The purpose of using external catalogues is to check the validity of our pipeline--cosmological constraints that are consistent with the mock cosmology suggests correct implementations of our algorithms. Their mock simulations use a flat $\Lambda$CDM cosmological model with the following cosmological parameters: $h=0.7$, $\Omega_\mathrm{m}=0.279$, $\Omega_\mathrm{b}=0.046$, $\sigma_8=0.82$, $n_\mathrm{s}=0.97$.

To generate the mock catalogues, we first download the first four realisations of the ray-tracing simulations with $N_\mathrm{side}=8192$, which is the highest resolution with multiple realisations available. For each mock catalogue realisation, we randomly crop 19 $3\times3\unit{{deg}^2}$ sub-fields from a ray-tracing realisation and extract shears $\boldsymbol{\gamma}\equiv(\gamma_1,\gamma_2)$ and convergences $\kappa$ at the positions $\boldsymbol{x}\equiv(x,y)$  and redshifts $z$ of the galaxies. Since the snapshots in the ray-tracing simulations from T17 are taken at discrete redshifts, we linearly interpolate between adjacent snapshots to get the shears and convergences at arbitrary redshifts, i.e.
\begin{gather}
\kappa(\boldsymbol{x},z)=\frac{z_{i+1}-z}{z_{i+1}-z_i}\kappa(\boldsymbol{x},z_i)+\frac{z-z_i}{z_{i+1}-z_i}\kappa(\boldsymbol{x},z_{i+1}), \\
\boldsymbol{\gamma}(\boldsymbol{x},z)=\frac{z_{i+1}-z}{z_{i+1}-z_i}\boldsymbol{\gamma}(\boldsymbol{x},z_i)+\frac{z-z_i}{z_{i+1}-z_i}\boldsymbol{\gamma}(\boldsymbol{x},z_{i+1}),
\end{gather}
where $z_i$ denotes the redshift of the $i^\text{th}$ snapshot, and $z_i\le z\le z_{i+1}$. This process of randomly cropping sub-fields and extracting shears is repeated ten times on each ray-tracing simulation, making 40 T17 mock catalogue realisations in total.

\section{Simulation pipeline}
\label{sec:simulation}

\begin{figure}[!t]
\centering
\includegraphics[width=6.5cm]{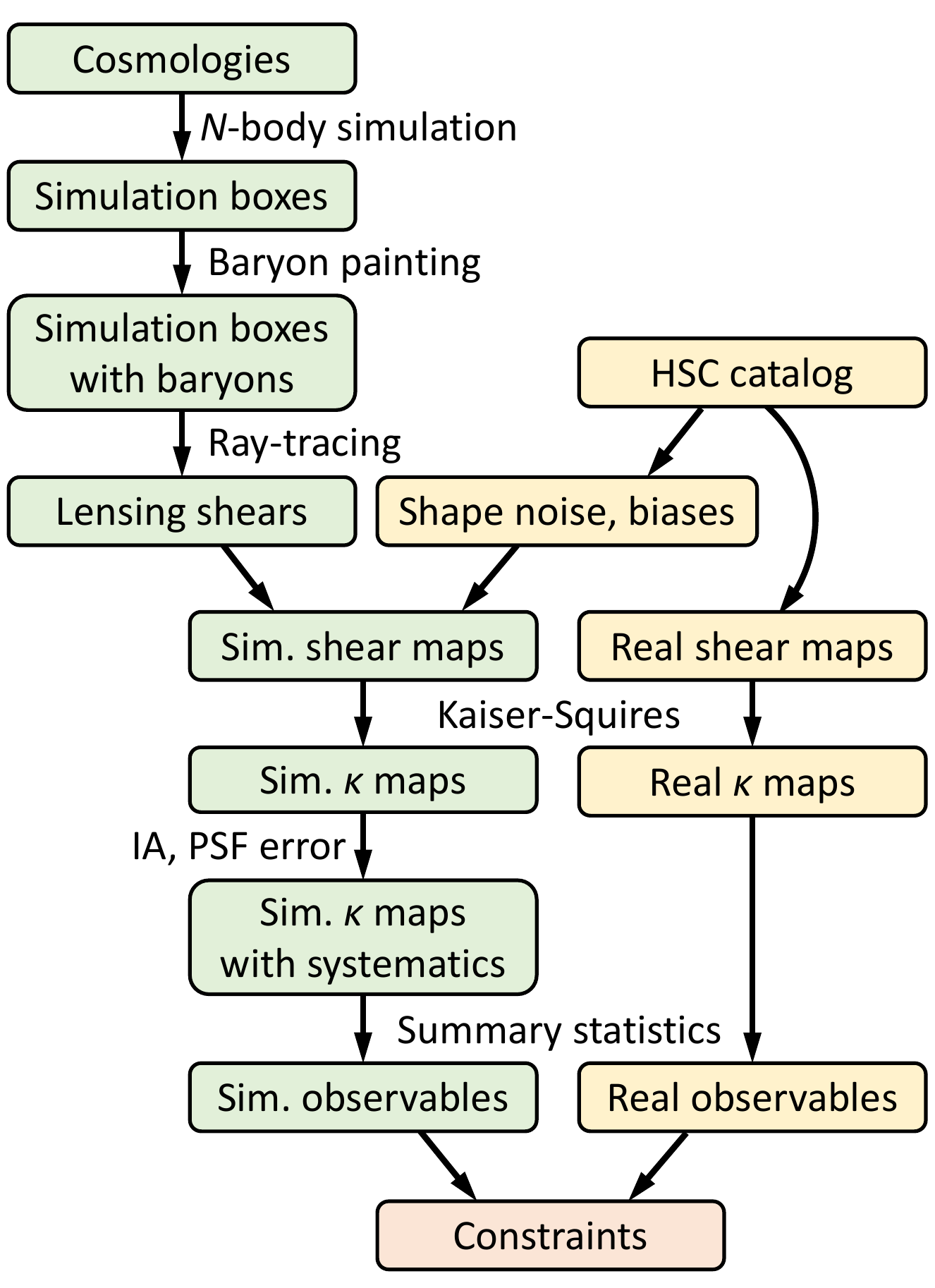}
\caption{An overview of our simulation pipeline (green). Some of the operations are also used in processing real data (yellow).
}
\label{fig:pipeline}
\end{figure}

Conceptually, our simulation pipeline is a forward modelling process that takes a set of parameters describing the cosmology, baryonic model, and systematics, and yields random realisations of weak lensing $\kappa$ maps. The simulated maps have similar statistical properties as the maps from the real data, so by varying the input cosmological parameters, the distribution of observables from the simulated maps can be used to constrain the cosmological model of the real Universe. In addition, we use those simulated maps to train deep learning models that can predict their underlying parameters, and the trained models can be applied to either another set of simulated maps or the real maps from the data, in the same way.

Figure~\ref{fig:pipeline} shows the flowchart diagram of our analysis pipeline, and in this section, we describe each of these steps in detail.

\subsection{\textit{N}-body simulation setup}
\label{sec:nbody}

\begin{figure}[!t]
\centering
\includegraphics[width=7.5cm]{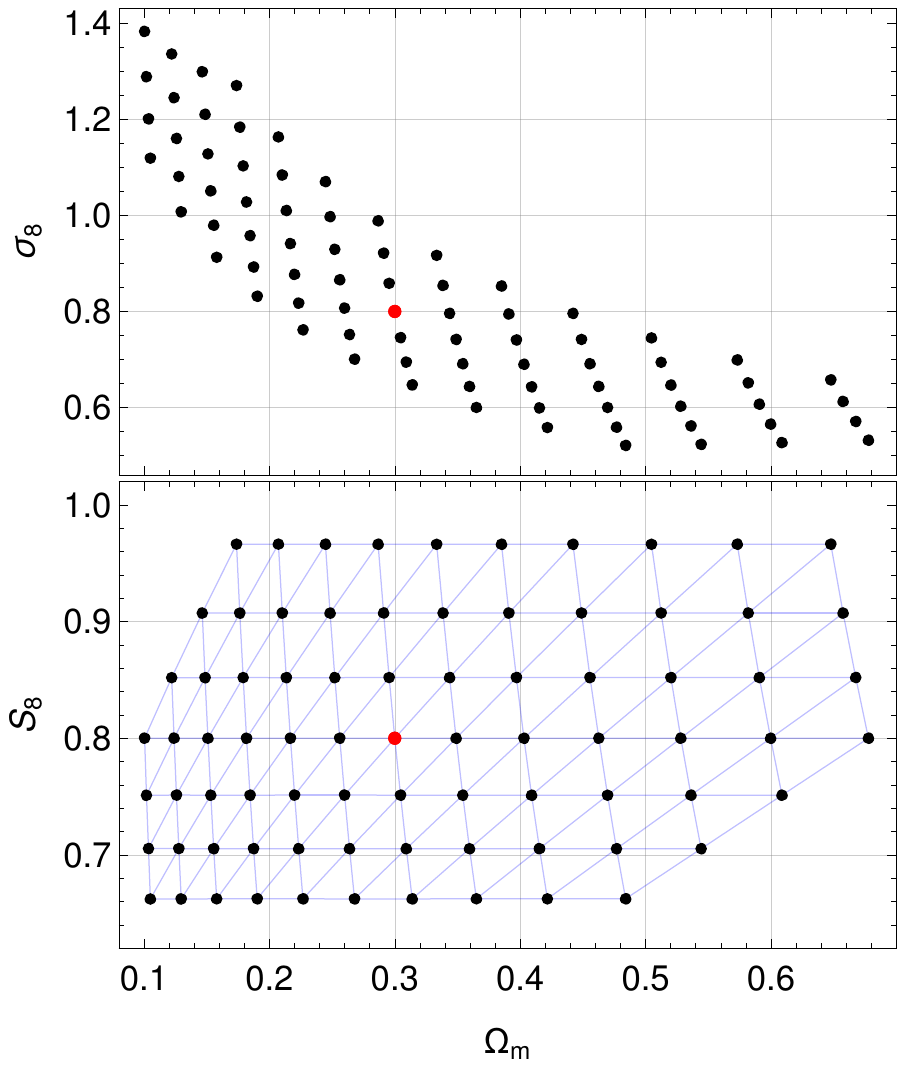}
\caption{The cosmological parameters $(\Om,\s8)$ in our suite of $N$-body simulations, where $S_8$ is defined as $(\Om/0.3)^{0.5}\s8$. The red dot denotes the fiducial cosmology $\Om=0.3, \s8=0.8$. The light blue triangle pattern shows the Delaunay triangulation of the grid used for interpolation.}
\label{fig:cosmologies}
\end{figure}

Our simulation suite consists of 79 DM-only $N$-body simulations with flat $\Lambda$CDM cosmology. Each simulation has its unique pair of $(\Om,\s8)$, with the rest of the cosmological parameters taken to be the marginalised means from the Planck 2018 results (TT,TE,EE+lowE+lensing): $h_0=0.6736$, $n_\mathrm{s}=0.9649$, and $\Ob=0.0493$ \citep{planck2018}. Our choices for $\Om$ and $\s8$, shown in Figure~\ref{fig:cosmologies}, are placed along constant $S_8$ values, which is close to the best-constrained parameter combination in most weak lensing analyses. The range $0.662\le S_8\le0.966$ is chosen to cover the posterior distributions from recent surveys including HSC Y1 \citep{hikage2019,hamana2020}, DES Y1 \citep{troxel2018}, and KiDS-1000 \citep{asgari2021}.

For each set of cosmological parameters, we use \textsc{music} \citep{hahn2011} to generate the initial condition of the DM-only simulation at $z_\mathrm{start}=50$, with a $500\hMpc$ box size and $1024^3$ particles. Then, we use \textsc{pkdgrav3} \citep{potter2017} to evolve the simulation box from $z=50$ to $z=0$ with 500 time steps. We choose the default schedule for the opening angle parameter: $\theta=0.4$ at $z>20$, $\theta=0.55$ at $2<z<20$, and $\theta=0.7$ at $z<2$, where large $\theta$ means higher force accuracy. We refer the reader to \citet{potter2017} for a detailed explanation.

\begin{table*}[!t]
\begin{tabular}{ccc|ccc|ccc}
\hline
Index & $z_\mathrm{snap}$ & $\chi_\mathrm{fid}$ & Index & $z_\mathrm{snap}$ & $\chi_\mathrm{fid}$ & Index & $z_\mathrm{snap}$ & $\chi_\mathrm{fid}$ \\
 &  & [$\hMpc$] &  &  & [$\hMpc$] &  &  & [$\hMpc$] \\ \hline
1 & 0.020 & 60 & 13 & 0.579 & 1500 & 25 & 1.413 & 2940 \\
2 & 0.060 & 180 & 14 & 0.635 & 1620 & 26 & 1.504 & 3060 \\
3 & 0.102 & 300 & 15 & 0.693 & 1740 & 27 & 1.600 & 3180 \\
4 & 0.145 & 420 & 16 & 0.753 & 1860 & 28 & 1.700 & 3300 \\
5 & 0.188 & 540 & 17 & 0.815 & 1980 & 29 & 1.806 & 3420 \\
6 & 0.232 & 660 & 18 & 0.879 & 2100 & 30 & 1.917 & 3540 \\
7 & 0.278 & 780 & 19 & 0.946 & 2220 & 31 & 2.034 & 3660 \\
8 & 0.325 & 900 & 20 & 1.016 & 2340 & 32 & 2.158 & 3780 \\
9 & 0.373 & 1020 & 21 & 1.089 & 2460 & 33 & 2.290 & 3900 \\
10 & 0.422 & 1140 & 22 & 1.164 & 2580 & 34 & 2.429 & 4020 \\
11 & 0.473 & 1260 & 23 & 1.244 & 2700 & 35 & 2.576 & 4140 \\
12 & 0.525 & 1380 & 24 & 1.327 & 2820 & 36 & 2.733 & 4260 \\ \hline
\end{tabular}
\caption{\label{tab:snapshots}The indices, redshifts, and comoving distances (fiducial cosmology) of the snapshots.}
\end{table*}

We take 36 snapshots at fixed redshifts for all cosmologies between $z=0.020$ and $z=2.733$ as listed in Table~\ref{tab:snapshots}. The redshifts $z_\mathrm{snap}^{(i)}$ are selected such that their corresponding comoving distances at the fiducial cosmology ($\Om=0.3$, $\s8=0.8$) are
\begin{equation}
\chi_\mathrm{fid}^{(i)}=(i-0.5)\times120\hMpc\;(i=1,2,\cdots,36).
\end{equation}
We note that because the snapshots can only be taken at the end of time steps, the actual redshifts are lower than the values in Table~\ref{tab:snapshots}. But since the number of steps is large, the deviations are very small: $|\Delta z|\sim0.003$ and $|\Delta \chi|\sim5\hMpc$, which are negligible compared to the estimated systematic error of photo-$z$ (see Section~\ref{sec:raytracing}).

\subsection{Ray-tracing}
\label{sec:raytracing}

\begin{figure}[!t]
\centering
\includegraphics[width=8.2cm]{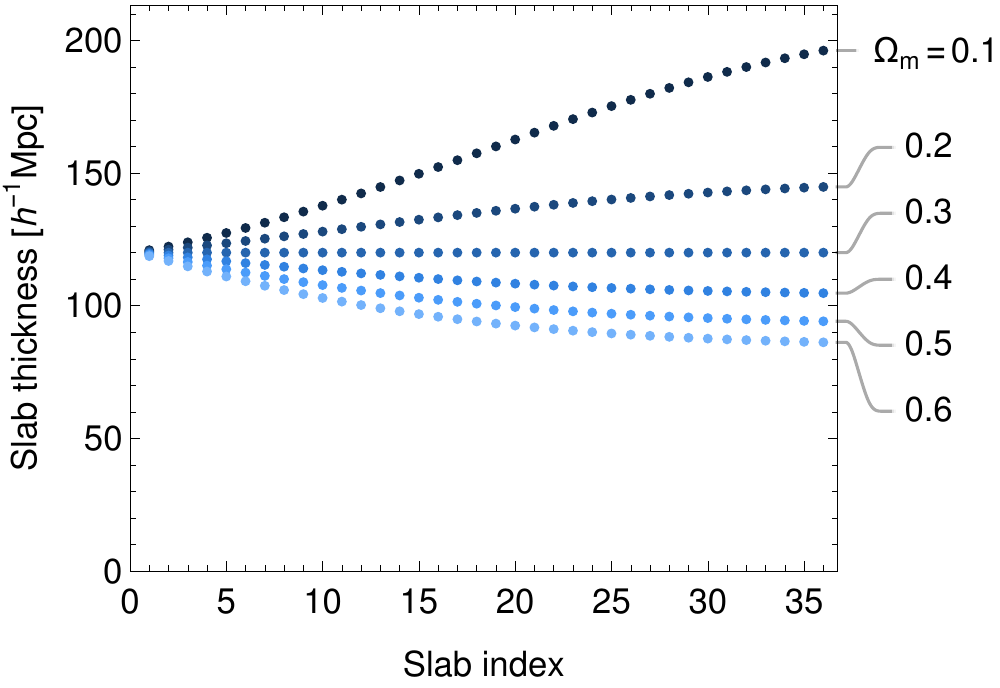}
\caption{The thickness of the slab as a function of the slab index and $\Om$. When $\Om=0.3$, all slabs have a constant thickness of $120\hMpc$.}
\label{fig:slab-thickness}
\end{figure}

In the real Universe, the shape of a galaxy at $z_\mathrm{s}$ is distorted by the foreground matter distribution between $z=0$ and $z=z_\mathrm{s}$ due to general relativity.  \emph{Ray-tracing} reproduces this process by tracing the light through the matter distribution from an $N$-body simulation. Ray-tracing is necessary for cosmological inference from small angular scales, using beyond-Gaussian statistics~\citep{petri2017}. The multi-lens-plane algorithm \citep{jain2000b, hilbert2009} is a popular implementation of ray-tracing. In this algorithm, the foreground matter distribution is discretised into multiple density slabs, which are taken from simulation snapshots. Then, each slab is viewed as a thin plane located at the centre of its thickness. When the light rays are traced from the observer at $z=0$ to the source galaxies, they only bend at the locations of those planes.

Given the redshifts of the snapshots $z_\mathrm{snap}^{(i)}$, we can calculate the desired thickness of the $i^\text{th}$ slab by
\begin{equation}
\Delta\chi^{(i)}=\frac{1}{2}\left[\chi_\mathrm{snap}^{(i+1)}-\chi_\mathrm{snap}^{(i-1)}\right],
\end{equation}
where $\chi_\mathrm{snap}^{(i)}$ denotes the comoving distance at $z_\mathrm{snap}^{(i)}$. We note that the thickness of the slabs is dependent on the cosmology: for the fiducial cosmology, the slabs have a constant thickness of $120\hMpc$ across all redshifts due to the snapshots being equally spaced in comoving distance, while for other cosmologies, the thickness at high redshifts can vary from $80\hMpc$ to $200\hMpc$, as shown in Figure~\ref{fig:slab-thickness}. We cut two to six slabs from each snapshots along each of the three axes, depending on the slab thickness. Then, we generate density planes from the slabs by calculating the column density on a $8192\times8192$ grid, followed by generating potential planes through solving the two-dimensional Poisson equation.

Our implementation of the ray-tracing algorithm strictly follows that presented by \citet{petri2013}. Each ray-tracing run starts from creating a stack of potential planes, where the plane at each redshift is randomly chosen, flipped, rotated (in multiples of $90^\circ$), and translated. We trace the light ray from the observer towards the direction of each galaxy until its ray-tracing redshift $z_\mathrm{rt}$ and calculate the total distortion in terms of lensing shears $\boldsymbol{\gamma}$ and the convergence $\kappa$.

\subsection{Redshift distribution uncertainty}
\label{sec:redshift-uncertainty}

In Section~\ref{sec:hsc-catalog}, we have estimated the redshift distribution $p_b(z)$ using the COSMOS reweighting method and assigned the ray-tracing redshifts $z_\mathrm{rt}$ to the galaxies, which are used as their source redshifts in \S~\ref{sec:raytracing}. While this method addresses the uncertainty of individual redshifts, the uncertainty of the whole distribution can remain a problem. In Figure~\ref{fig:z-rt}, one can find systematic differences between \texttt{ephor\_AB} $z_\mathrm{best}$ and $z_\mathrm{rt}$, suggesting that $p_b(z)$ may be systematically biased along the redshift axis due to our choice of the photo-$z$ algorithm. In addition, the discrepancies in the average $z_\mathrm{best}$ among various photo-$z$ algorithm, even trained with the same COSMOS 30-band redshift, suggest that photo-$z$ algorithms are associated with intrinsic statistical errors \citep{tanaka2018,hikage2019,hamana2020}.

In this work, we follow \citet{hikage2019} to model the uncertainty in each redshift distribution. The uncertainty of $p_b(z)$ is parameterised with a translation by $\Delta z_b$:
\begin{equation}
\tilde{p}_b(z) = p_b(z+\Delta z_b),
\end{equation}
and hence,
\begin{equation}
\tilde{z}_\mathrm{rt} = z_\mathrm{rt}+\Delta z_b,
\end{equation}
where $\tilde{p}_b(z)$ denotes the redshift distribution without the overall redshift error, and $\tilde{z}_\mathrm{rt}$ the ray-tracing redshift sampled from $\tilde{p}_b(z)$. The priors of $\Delta z_b$ follow normal distributions $\mathcal{N}(0,\sigma_{\Delta z_b})$, and \citet{hikage2019} have estimated the values of $\sigma_{\Delta z_b}$ to be: 0.0285 ($0.3<z<0.6$), 0.0135 ($0.6<z<0.9$), 0.0383 ($0.9<z<1.2$), and 0.0376 ($1.2<z<1.5$), shown in Figure~\ref{fig:z-rt}. \citet{hamana2020} have shown that the two-point correlation function of the HSC first-year catalogue cannot provide better constraints on $\Delta z_b$ than their priors. Therefore, we do not include $\Delta z_b$ as free parameters, but instead we simply add randomly sampled $\Delta z_b$ values to $z_\mathrm{rt}$ during ray-tracing---effectively marginalising over the priors of $\Delta z_b$.

\subsection{Baryonic model}
\label{sec:baryonic-model}

\begin{table*}[!t]
\centering
\begin{tabular}{llll}
\hline
Parameter & Fiducial value & Prior \\ \hline
\multicolumn{3}{l}{Cosmological and baryonic parameters} \\
$\Om$ & $0.3$ & flat within the boundary of simulated cosmologies \\
$\s8$ & $0.8$ & flat within the boundary of simulated cosmologies \\
$A_\mathrm{IA}$ & $0$ & flat $[-3,3]$ \\
$M_\mathrm{c}$ & $3.3\ee{13}h^{-1}\Msun$ & log-uniform $[10^{12}h^{-1}\Msun,10^{16}h^{-1}\Msun]$ \\
$M_{1,0}$ & $8.63\ee{11}h^{-1}\Msun$ & log-uniform $[10^{10}h^{-1}\Msun,10^{13}h^{-1}\Msun]$ \\
$\eta$ & $0.54$ & log-uniform $[10^{-0.7},10^{0.5}]$ \\
$\beta$ & $0.12$ & log-uniform $[10^{-1.0},10^{0.5}]$ \\ \hline
\multicolumn{3}{l}{Nuisance parameters} \\
$\Delta z_1$ & $-$ & Gaussian $\mathcal{N}(0,0.0285)$ \\
$\Delta z_2$ & $-$ & Gaussian $\mathcal{N}(0,0.0135)$ \\
$\Delta z_3$ & $-$ & Gaussian $\mathcal{N}(0,0.0383)$ \\
$\Delta z_4$ & $-$ & Gaussian $\mathcal{N}(0,0.0376)$ \\
$\Delta m$ & $-$ & Gaussian $\mathcal{N}(0,0.01)$ \\
$\alpha_\mathrm{psf}$ & $-$ & Gaussian $\mathcal{N}(0.030,0.015)$ \\
$\beta_\mathrm{psf}$ & $-$ & Gaussian $\mathcal{N}(-0.89,0.70)$ \\ \hline
\end{tabular}
\caption{The prior distributions of all parameters.}
\label{tab:prior}
\end{table*}

We employ the baryonic correction model \citep[BCM,][see also \citealt{schneider2015}]{arico2020} to make modifications to the simulated density fields, mimicking the impact of cooling, star formation, and feedback during galaxy formation. Other than the baryon density parameter $\Ob$, the BCM has four free parameters: two characteristic halo masses $M_\mathrm{c}$ and $M_{1,0}$, the maximum distance $\eta$ of the gas ejected from halos, and the logarithmic slope $\beta$ of the gas fraction versus halo mass. With the total mass of each halo conserved, these four parameters control the difference between the density profile with and without baryons.

The implementation of the BCM in this work is identical to our previous works \citep[see][]{lu2021, lu2022}. We briefly introduce the process here, and we refer the reader to \citet{lu2021} for a detailed description. In the analyses where the baryonic effects are included, the BCM is applied to the density planes prior to ray-tracing. In each halo, a fixed portion ($\Ob/\Om$) of the mass is replaced by the density profile calculated with the halo mass, concentration parameter, and baryonic parameters. The fiducial values of the baryonic parameters are: $M_\mathrm{c}=3.3\ee{13}h^{-1}\Msun$, $M_{1,0}=8.63\ee{11}h^{-1}\Msun$, $\eta=0.54$, $\beta=0.12$, and we adopt log-uniform priors for all four parameters with the ranges shown in Table~\ref{tab:prior}. Note that the prior ranges in this work are different from those in \citet{arico2020}, because we use the constraints by other observations \citep[see e.g.][]{schneider2015, lu2021} as reference instead of using the best-fits from hydrodynamical simulations.

\subsection{Simulated ellipticity}
\label{sec:simulated-ellipticity}

A measured galaxy ellipticity presented in the real shear catalogue is the consequence of multiple factors, with the lensing shears calculated in
\S~\ref{sec:raytracing} being only one of them. In this section, we present our method of simulating ellipticities considering the other two major factors---shape noise and shear measurement biases, as part of the forward modelling process.

Shape noise refers to part of the ellipticity that is contributed by the intrinsic shape of the galaxy, ignoring the correlation between the intrinsic shapes. In terms of generating shape noise to be added to the simulated shears, a simple empirical method is randomly rotating the measured ellipticity $\boldsymbol{e}\equiv(e_1,e_2)$. This ensures that (1) the level of shape noise is almost the same as that in the real catalogue (because $|\boldsymbol{\gamma}|\ll|\boldsymbol{e}|$ for a single galaxy) and (2) random rotations removes lensing signals from the ellipticities.

Shear measurement biases, specifically the multiplicative bias $m$ and the additive bias $\boldsymbol{c}\equiv(c_1,c_2)$, describe the relation between the true shear $\boldsymbol{\gamma}^\mathrm{(true)}$ and the measured shear $\boldsymbol{\gamma}^\mathrm{(mea)}$:
\begin{equation}
\boldsymbol{\gamma}^\mathrm{(mea)}=(1+m)\boldsymbol{\gamma}^\mathrm{(true)}+\boldsymbol{c}.
\end{equation}
There are multiple reasons for the existence of shear biases, such as the limitations of the PSF deconvolution algorithm, or the shape of the galaxy not being correctly described by the model. \citet{mandelbaum2018b} have done thorough investigations into shear biases using image simulations, and they provide $m$ and $\boldsymbol{c}$ estimations for each galaxy in the HSC shear catalogue. With all calibrated biases removed, \citet{mandelbaum2018b} show that the uncertainty of the residual multiplicative bias $\Delta m$ is controlled at the 1 percent level.

In our pipeline, the simulated ellipticity of a galaxy is calculated following \citet{bernstein2002} and \citet{shirasaki2019}, but we neglect the second order terms as we find virtually no difference in the produced maps:
\begin{gather}
\boldsymbol{e}^\mathrm{(sim)}=e^{i\phi}\boldsymbol{e}^\mathrm{(mea)}+(1+\Delta m)(1+m_\mathrm{tot})\frac{2\boldsymbol{\gamma}\mathcal{R}}{1-\kappa}+\boldsymbol{c}, \label{eqn:esim}\\
\mathcal{R}\equiv 1-{\frac{1}{2}\left|\boldsymbol{e}^\mathrm{(mea)}\right|}^2\frac{e_\mathrm{RMS}^2}{e_\mathrm{RMS}^2+\sigma_\mathrm{e}^2}, \\
m_\mathrm{tot}\equiv m+m_\mathrm{sel}+m_\mathcal{R},
\end{gather}
where $\boldsymbol{e}^\mathrm{(mea)}$ denotes the measured ellipticity in the catalogue, $e^{i\phi}\boldsymbol{e}^\mathrm{(mea)}$ the measured ellipticity with a random rotation, $m$ and $\boldsymbol{c}$ the estimated biases in the HSC catalogue, $e_\mathrm{RMS}$ the estimated intrinsic root mean square (RMS) ellipticity in the catalogue, and $\sigma_\mathrm{e}$ the estimated measurement error in the catalogue. Following \citet{hikage2019}, the multiplicative biases due to galaxy size selection $m_\mathrm{sel}$ are 0.0086, 0.0099, 0.0091, and 0.0091 in the four redshift bins respectively (low-$z$ to high-$z$), the multiplicative biases due to redshift-dependent responsivity corrections $m_\mathcal{R}$ are 0.000, 0.000, 0.015, and 0.030 in the four bins respectively, and the nuisance parameter $\Delta m$ is randomly sampled from the normal distribution $\mathcal{N}(0,0.01)$, applied across all simulated ellipticities. A realisation of the simulated ellipticities can be viewed as a catalogue that have similar characteristics as the real catalogue,  except for the underlying model parameters and a few systematics that need to be addressed after map conversion.

\subsection{Shear maps}
\label{sec:shear-maps}

\begin{figure*}[!t]
\centering
\includegraphics[width=15.5cm]{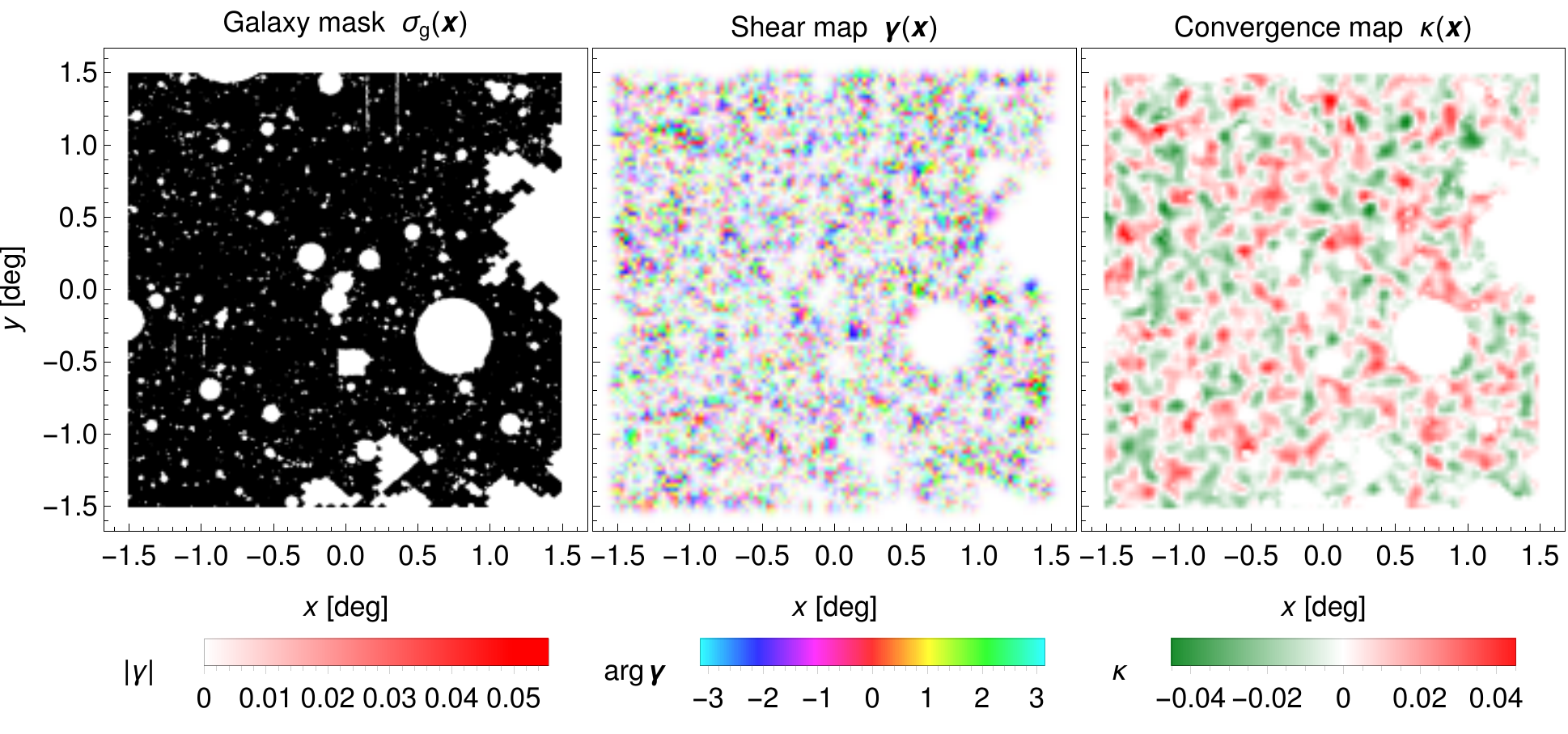}
\caption{The field mask, shear map, and convergence map of an HSC sub-field in GAMA09H. The shear map and convergence map are generated with a smoothing scale of $\sigma_\mathrm{G}=2\unit{arcmin}$.}
\label{fig:example-maps}
\end{figure*}

With the ellipticities simulated for all galaxies in a sub-field and a redshift bin, we can generate a $3\times3\unit{{deg}^2}$ shear map on a $104\times104$ square grid, and we add 12 pixels of padding on all four sides of the map to better maintain the lensing signal near the edge during smoothing (see below). The size of the full map is therefore $3.69\times3.69\unit{{deg}^2}$ with a dimension of $128\times128$, and the size of each pixel is $\Delta x=1.73\unit{arcmin}$.

In the HSC catalogue, there are regions without galaxies due to bright stars, and we need to find this \emph{field mask} $\sigma_\mathrm{mask}(\boldsymbol{x})$ for each sub-field, which will be used in the generation of shear maps. Note that the field mask does not depend on the redshift bin, thus the number density of the galaxies is high enough for us to calculate the mask at a resolution of $0.87\unit{arcmin}$, i.e. on a $208\times208$ grid for a $3\times3\unit{{deg}^2}$ sub-field. The sub-fields have 10 to 15 galaxies per pixel on average among all pixels with at least one galaxy, and based on that, we consider the pixels with fewer than four galaxies to be covered by the mask ($\sigma_\mathrm{mask}(\boldsymbol{x})=1$). Then, the field masks are down-sampled by a factor of two to the resolution of shear maps, an example of which is shown in Figure~\ref{fig:example-maps}.

The calculation of the shear maps follows the shear estimation procedure as in Appendix 3 of \citet{mandelbaum2018a}. The value of the pixel centred at $\boldsymbol{x}\equiv(x,y)$ is
\begin{equation}
\boldsymbol{\gamma}(\boldsymbol{x})=\frac{1}{1+m_\mathrm{tot}(\boldsymbol{x})}\left[\frac{\boldsymbol{e}(\boldsymbol{x})}{2(1-e_\mathrm{RMS}^2(\boldsymbol{x}))}-\boldsymbol{c}(\boldsymbol{x})\right].
\end{equation}
The fields on the right-hand side are defined as
\begin{gather}
\phi(\boldsymbol{x})=\frac{\phi_w(\boldsymbol{x})}{w(\boldsymbol{x})}, \label{eqn:field-phi}\\
\phi_w(\boldsymbol{x})=\left(\sum_{\boldsymbol{x}_i\in A(\boldsymbol{x})}{w_i\phi_i}\right)+\sigma_\mathrm{mask}(\boldsymbol{x})\langle w\rangle\langle\phi\rangle, \label{eqn:field-phiw}\\
w(\boldsymbol{x})=\left(\sum_{\boldsymbol{x}_i\in A(\boldsymbol{x})}{w_i}\right)+\sigma_\mathrm{mask}(\boldsymbol{x})\langle w\rangle, \label{eqn:field-w}\\
\langle w\rangle=\frac{1}{N}\sum_{i}{w_i},\;\langle\phi\rangle=\frac{1}{N}\sum_{i}{\phi_i},
\end{gather}
where $\phi$ is one of $\{e_\mathrm{RMS}^2, m_\mathrm{tot}, \boldsymbol{c}, \boldsymbol{e}\}$, $N$ the number of galaxies in the sub-field and the redshift bin, $w_i$ the lensing weight of the $i^\text{th}$ galaxy in the catalogue, and $\boldsymbol{x}_i\in A(\boldsymbol{x})$ means that the $i^\text{th}$ galaxy is located in the pixel centred at $\boldsymbol{x}$. We further simplify the calculation by assuming $\langle\boldsymbol{c}\rangle=\langle\boldsymbol{e}\rangle=0$. Note that due to the second terms in Equation~\eqref{eqn:field-phiw} and \eqref{eqn:field-w}, the average values (e.g. shears and biases) are assigned to the masked region of the map so that $\phi(\boldsymbol{x})$ and $\boldsymbol{\gamma}(\boldsymbol{x})$ are well-defined on the entire map. When smoothing of the shear map is needed, a Gaussian kernel with standard deviation $\sigma_\mathrm{G}$ is applied to $\phi_w(\boldsymbol{x})$ and $w(\boldsymbol{x})$, e.g.
\begin{equation}
w(\boldsymbol{x}_0)\rightarrow\iint\mathrm{d}\boldsymbol{x}\,\frac{\exp\left(-{|\boldsymbol{x}|}^2/2\sigma_\mathrm{G}^2\right)}{\sqrt{2\pi\sigma_\mathrm{G}}}w(\boldsymbol{x}_0+\boldsymbol{x}).
\end{equation}

The process of shear map generation from the real HSC catalogue is identical to the that from the simulated catalogues. An example shear map from the real HSC data with smoothing is shown in Figure~\ref{fig:example-maps}.

\subsection{Convergence maps}
\label{sec:convergence-maps}

We use the Kaiser--Squires method \citep{kaiser1993} to convert the shear maps from the previous step to convergence maps $\kappa(\boldsymbol{x})$. In Fourier space, the shear $\hat{\boldsymbol{\gamma}}(\boldsymbol\ell)$ and the convergence $\hat{\kappa}(\boldsymbol\ell)$ are related by
\begin{equation}
\hat{\kappa}(\boldsymbol\ell)=\frac{\ell_1^2-\ell_2^2}{\ell_1^2+\ell_2^2}\hat{\gamma_1}(\boldsymbol\ell)+\frac{2\ell_1 \ell_2}{\ell_1^2+\ell_2^2}\hat{\gamma_2}(\boldsymbol\ell).
\end{equation}
Again, the same conversion is applied to the real HSC shear maps. An example convergence map from the HSC data is shown in Figure~\ref{fig:example-maps}. All of our cosmological analyses will be done convergence maps.

\subsection{Intrinsic alignment}
\label{sec:ia}

The lensing shear from the ray-tracing method in
\S~\ref{sec:raytracing} reproduces the interaction between the light from the source galaxy and the foreground matter distribution and the intrinsic ellipticities of the galaxies are randomly assigned. But in the real Universe, there can be an intrinsic alignment (IA) between the galaxies through two mechanisms: (1) galaxies that are physically close to each other tend to have similar ellipticities (intrinsic-intrinsic, or II), and (2) a foreground galaxy is correlated with a local matter distribution that distorts a background galaxy in the orthogonal direction (gravitational-intrinsic, or GI). In this work, we adopt the non-linear alignment model \citep{hirata2004,bridle2007} which predicts the power spectrum of the IA signal to be
\begin{eqnarray}
\nonumber P_\mathrm{IA}(k,z)&=&P_\mathrm{II}(k,z)+P_\mathrm{GI}(k,z)\\
&=&F^2(z) P_\delta(k,z) + F(z) P_\delta(k,z),
\end{eqnarray}
where
\begin{equation}
F(z)=-A_\mathrm{IA}C_1\bar{\rho}(z)\frac{D(z)}{D(0)}\left(\frac{1+z}{1+z_0}\right)^\eta\left(\frac{\bar{L}}{L_0}\right)^\beta.
\label{eqn:fz}
\end{equation}
In Equation~\eqref{eqn:fz}, $A_\mathrm{IA}$ denotes a free parameter that controls the amplitude of IA, $C_1=5\ee{-14}h^{-2}\Msun^{-1}\mathrm{Mpc}^{3}$ a normalising constant, $\bar{\rho}(z)$ the mean matter density of the universe at redshift $z$, $D(z)$ the linear growth factor, and $\bar{L}$ the weighted average luminosity of the source sample. Following \citet{fluri2019} and \citet{hildebrandt2017}, we ignore the redshift and luminosity dependent terms by setting $\eta=\beta=0$.

A natural simple choice for the IA-induced convergence field is linked to the matter distribution. Specifically, we follow \citet{fluri2019} and adopt 
\begin{equation}
\kappa_\mathrm{IA}^{(b)}(\boldsymbol{x})=\int\mathrm{d}z\,F(z)n_\mathrm{g}^{(b)}(z)\delta(\boldsymbol{x},z),
\end{equation}
where $n_\mathrm{g}^{(b)}(z)$ denotes the number of galaxies per redshift in the bin $b$ normalised with $\int\mathrm{d}z\,n_\mathrm{g}^{(b)}(z)=1$, and $\delta(\boldsymbol{x},\chi)$ the density contrast. This IA estimation yields the correct IA power spectrum by construction and can be easily incorporated into our ray-tracing subroutine. We can trace $\kappa_\mathrm{IA}(\boldsymbol{x})$ along with $\boldsymbol{\gamma}_\mathrm{IA}(\boldsymbol{x})$ for each galaxy. We note that $\kappa_\mathrm{IA}$ and $\boldsymbol{\gamma}_\mathrm{IA}$ are linearly proportional to $A_\mathrm{IA}$,  and thus we only need to trace them with $A_\mathrm{IA}=1$ to get $\tilde{\kappa}_\mathrm{IA}(\boldsymbol{x})$ and $\tilde{\boldsymbol{\gamma}}_\mathrm{IA}(\boldsymbol{x})$---results for any other $A_\mathrm{IA}$ can be inferred via 
\begin{equation}
\kappa_\mathrm{IA}(\boldsymbol{x})=A_\mathrm{IA}\tilde{\kappa}_\mathrm{IA}(\boldsymbol{x}),\;\boldsymbol{\gamma}_\mathrm{IA}(\boldsymbol{x})=A_\mathrm{IA}\tilde{\boldsymbol{\gamma}}_\mathrm{IA}(\boldsymbol{x}).
\end{equation}

In our analyses where IA is included in the model, we modify the calculation of the simulated ellipticity in Equation~\eqref{eqn:esim} by doing the following replacement:
\begin{equation}
\frac{2\boldsymbol{\gamma}\mathcal{R}}{1-\kappa}\rightarrow\frac{2(\boldsymbol{\gamma}+A_\mathrm{IA}\tilde{\boldsymbol{\gamma}}_\mathrm{IA})\mathcal{R}}{1-(\kappa+A_\mathrm{IA}\tilde{\kappa}_\mathrm{IA})}.
\end{equation}

\subsection{PSF error}
\label{sec:psf-error}

\begin{figure}[!t]
\centering
\includegraphics[width=7.5cm]{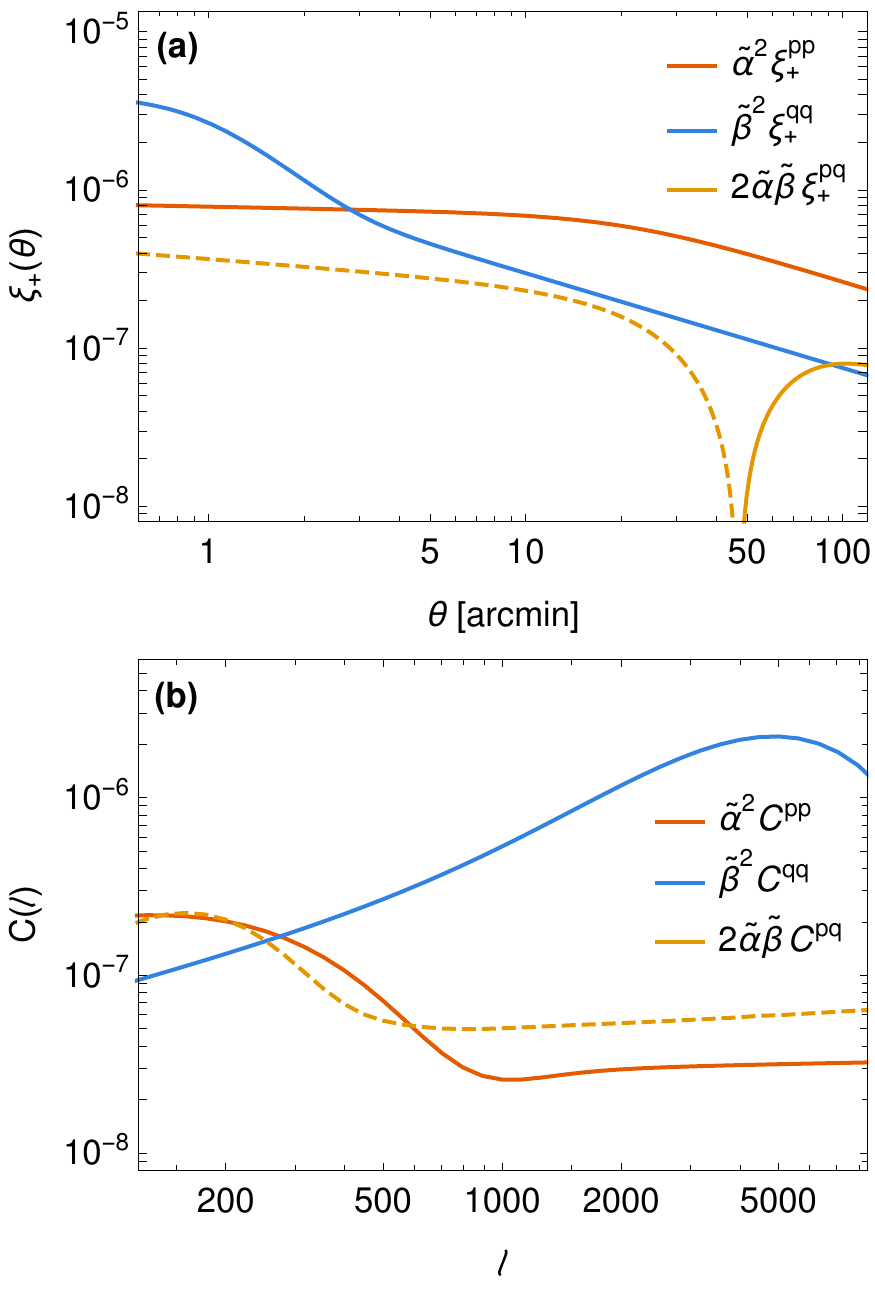}
\caption{(a) The auto and cross-correlation functions and (b) the auto and cross power spectra of the PSF residual terms, where $\tilde{\alpha}_\mathrm{psf}=0.030$ and $\tilde{\beta}_\mathrm{psf}=-0.89$. Dashed lines mean that the values are negative.}
\label{fig:psf-error}
\end{figure}

The shape of a galaxy when observed by a ground-based telescope is contaminated by effects like atmospheric disturbances and instrument characteristics. They can be modelled by a PSF, which modifies the shape of the source through convolution. A typical approach to remove the PSF from a galaxy shape involves two steps: (1) predict the shape of the PSF at the location of the galaxy using nearby stellar images, and (2) apply a PSF deconvolution algorithm, such as the re-Gaussianization method in the HSC pipeline. Nevertheless, both steps can still leave small residual PSF in the galaxy shapes: (1) the number density of stars is much smaller than that of galaxies, thus the variations of PSF on small angular scales can not be accurately predicted, and (2) the PSF deconvolution algorithm can have imperfections that leaks PSF shape into the deconvolved galaxy shape \citep{mandelbaum2018a}. We follow \citet{hikage2019} and \citet{hamana2020} to model PSF residuals as a linear combination of these two effects:
\begin{equation}
\delta\boldsymbol{\gamma}=\alpha_\mathrm{psf}\boldsymbol{\gamma}^\mathrm{p}+\beta_\mathrm{psf}\boldsymbol{\gamma}^\mathrm{q},
\end{equation}
where $\boldsymbol{\gamma}^\mathrm{p}=\boldsymbol{e}^\mathrm{psf}/2$ denotes the shear associated with the predicted PSF shape, and $\boldsymbol{\gamma}^\mathrm{q}=\boldsymbol{\gamma}^\mathrm{p}-\boldsymbol{\gamma}_\mathrm{true}^\mathrm{p}$ the difference between the predicted PSF shear and the untraceable true PSF shear.

As the small scale variations of PSFs is not recoverable, it is not possible to reproduce $\boldsymbol{\gamma}^\mathrm{q}$ fields that have identical characteristics to the real data. We therefore adopt an alternative approach where we model the the PSF residuals as Gaussian random fields, and they can be generated based on the auto- and cross-correlation functions between $\boldsymbol{\gamma}^\mathrm{p}$ and $\boldsymbol{\gamma}^\mathrm{q}$ measured by \citet{hamana2020}. We choose the correlation functions because they are measured at a higher angular scale resolution than the power spectra by \cite{hikage2019}. We fit analytical models to the correlation functions presented in Figure~20 of \citet{hamana2020} and derive their power spectra counterparts ($C^\mathrm{pp}$, $C^\mathrm{pq}$, and $C^\mathrm{qq}$) by the Hankel transform
\begin{equation}
C^{ij}(\ell)=2\pi\int\mathrm{d}\theta\,\theta\,\xi_+^{ij}(\theta)J_0(\ell\theta).
\end{equation}
Visualisations of the correlation functions and power spectra are shown in Figure~\ref{fig:psf-error}. The total PSF error power spectrum is a linear combination of the three components:
\begin{equation}
C^\mathrm{psf}(\ell)=\alpha_\mathrm{psf}^2 C^\mathrm{pp}(\ell)+2\alpha_\mathrm{psf}\beta_\mathrm{psf} C^\mathrm{pq}(\ell)+\beta_\mathrm{psf}^2 C^\mathrm{qq}(\ell),
\end{equation}
where the nuisance parameters $\alpha$ and $\beta$ have Gaussian priors derived from Figure~21 in \citet{hamana2020} with all angular scales:
\begin{equation}
\alpha_\mathrm{psf}\sim\mathcal{N}(0.030,0.015), \;\beta_\mathrm{psf}\sim\mathcal{N}(-0.89,0.70).
\end{equation}
After the convergence maps are generated for a realisation (see Section~\ref{sec:convergence-maps}), we sample $\alpha_\mathrm{psf}$ and $\beta_\mathrm{psf}$ from their prior distributions and generate one PSF convergence map for each sub-field, which is added to the convergence maps for all redshift bins.

\section{Summary statistics}
\label{sec:statistics}

\subsection{Convolutional neural network}
\label{sec:cnn}

\subsubsection{Overview}
\label{sec:cnn-overview}

Considering that our simulation pipeline is a function that generates maps from parameters:
\begin{equation}
\textrm{pipeline: }(\boldsymbol{\theta},\boldsymbol{\nu},\boldsymbol{r})\rightarrow\kappa,
\end{equation}
where $\boldsymbol{\theta}\equiv(p_1,\cdots,p_N)$ denotes the parameters of interest, $\boldsymbol{\nu}$ the nuisance parameters, and $\boldsymbol{r}$ the random variables, we train a \emph{summarising CNN} to be a function that predicts the parameters of interest given a map:
\begin{equation}
\textrm{CNN: }\kappa\rightarrow\boldsymbol{y},
\end{equation}
where the weights of the CNN are adjusted so that $\boldsymbol{y}$ is close to $\boldsymbol{\theta}$, or more rigorously, minimising a \emph{loss function} $L(\boldsymbol{y},\boldsymbol{\theta})$. Here, we note that although the CNN is trained with the goal of reproducing the parameters of the input map by its outputs, those outputs usually come with large biases \citep[see e.g.][]{gupta2018, ribli2019,lu2022}, so they should not be used as estimations of those parameters. Instead, we view the output of the CNN as another summary statistic, and we derive a likelihood function $\mathcal{L}(\boldsymbol{y}|\boldsymbol{\theta})$ from the predictions regardless of the meaning of its outputs, effectively marginalising over the nuisance parameters and randomness. A neural network-based summary statistic works in a similar way to traditional statistics such as power spectra and peak counts in that they all take a convergence map (or maps) as the input and generate an array of numbers as the output. The difference is that a neural network has millions of free parameters (called ``weights'') that need to be determined through a training process. Using the CNNs in this way has been investigated in multiple studies \citep[see][]{gupta2018,ribli2019,fluri2019,fluri2022,lu2022}.

The CNN is often trained on a data set with a finite number of maps for multiple rounds. It is possible that the network over-fits to the training set, i.e. learning some irrelevant features out of randomness which only work on the training set. So it is important that we use a different set of maps, which have never been seen by the network, for the derivation of the likelihood function.

\subsubsection{Map smoothing}
\label{sec:cnn-smoothing}

The weak lensing signals on very small angular scales are often discarded for a few reasons: (1) the modelling of baryonic effects is not well-known, (2) the non-linear alignment model underestimates the IA signal on small scales, (3) the PSF modelling error becomes comparable to the lensing signal at low redshifts, and (4) very small scales are not very informative due to shape noise \citep{fang2007,singh2015,hikage2019,arico2020}. For example, \citet{hikage2019} adopts $\ell_\mathrm{max}=1900$ in their fiducial analysis ($\theta_\mathrm{min}\sim6\unit{arcmin}$), and \citet{hamana2020} adopts $\theta_\mathrm{min}=7\unit{arcmin}$ for $\xi_+(\theta)$. To be consistent with these two studies, we apply smoothing with Gaussian filters to our simulated maps (see Section~\ref{sec:shear-maps}) so that the CNN will not learn from the fluctuations on small scales which are not well-understood. Our fiducial analysis uses a smoothing radius of $\sigma_\mathrm{G}=4\unit{arcmin}$, which suppress the power by 90 percent (i.e. 10 percent of the power remaining) at $\ell=1400$ and by 99 percent at $\ell=2000$, comparable to the fiducial analysis in \citet{hikage2019}. We will also perform analysis with other smoothing scales  $\sigma_\mathrm{G}=1\text{--}8\unit{arcmin}$ (99 percent suppression at $\ell=8000\text{--}1000$) to see how the statistical errors change.

\subsubsection{Architecture}
\label{sec:cnn-architecture}

\begin{table}[!t]
\centering
\begin{tabular}{cccc}
\hline
Layer/ & Kernel & Stride & Output    \\
block  & size   &        & dimension \\ \hline
(Input)     &            &   & $4\times104\times104$               \\
Convolution & $7\times7$ & 2 & $n_\mathrm{ch}\times52\times52$     \\

Residual block 1                  & $-$ & $-$ & $n_\mathrm{ch}\times52\times52$ \\
$\vdots$                          &  &  & \\
Residual block $n_\mathrm{block}$ & $-$ & $-$ & $n_\mathrm{ch}\times52\times52$ \\

Pooling     & $2\times2$ & 2   & $     n_\mathrm{ch} \times26\times26$ \\
Convolution & $3\times3$ & 1   & $( 2\,n_\mathrm{ch})\times24\times24$ \\
Pooling     & $2\times2$ & 2   & $( 2\,n_\mathrm{ch})\times12\times12$ \\
Convolution & $3\times3$ & 1   & $( 4\,n_\mathrm{ch})\times10\times10$ \\
Pooling     & $2\times2$ & 2   & $( 4\,n_\mathrm{ch})\times5\times5$   \\
Convolution & $3\times3$ & 1   & $( 8\,n_\mathrm{ch})\times3\times3$   \\
Pooling     & $3\times3$ & 2   & $( 8\,n_\mathrm{ch})\times1\times1$   \\
Linear      & $-$        & $-$ & $256$                                 \\
ReLU        & $-$        & $-$ & $256$                                 \\
Linear      & $-$        & $-$ & $N_\theta$                                 \\
\hline
\end{tabular}
\caption{The architecture of the CNNs. Here $n_\mathrm{ch}$ (number of channels) and $n_\mathrm{block}$ (number of residual blocks) are hyper-parameters. Convolution layers, except those in the residual blocks, are implicitly followed by batch normalisation layers and then ReLU layers.}
\label{tab:architecture}
\end{table}

Similar to our previous work \citep{lu2022}, we employ a CNN architecture with residual blocks \citep{he2016}, as is shown in Table~\ref{tab:architecture}. A residual block consists of a $3\times3$ convolution layer, a batch normalisation layer, a rectified linear (ReLU) layer, another $3\times3$ convolution layer, a batch normalisation layer, and a ReLU layer in series, with a skip connection that add the residual block input to the intermediate result prior to the last ReLU layer.

The input to the CNN is the convergence maps of the four redshift bins with the padding removed, thus it is an array with dimension $4\times104\times104$, representing a sub-field with $3\times3\unit{{deg}^2}$ in size. The output from the CNN is a list of $N_\theta$ numbers, where $N_\theta=3$ when we adopt a model with only cosmological parameters ($\Om$, $\s8$, $A_\mathrm{IA}$) being free, or $N_\theta=7$ when we adopt a model with four additional free baryonic parameters ($M_\mathrm{c}$, $M_{1,0}$, $\eta$, $\beta$). The network has two hyper-parameters: the number of map channels in residual blocks $n_\mathrm{ch}$ and the number of residual blocks $n_\mathrm{block}$. Our main models as presented in Section~\ref{sec:constraints} and Section~\ref{sec:discussion} uses $n_\mathrm{ch}=64$ and $n_\mathrm{block}=10$; models with other choices are discussed in Section~\ref{sec:hyperparameters}.

The loss function of the CNN is
\begin{equation}
L(\boldsymbol{y},\boldsymbol{\theta})=\frac{1}{N_\theta}\sum_{i=1}^{N_\theta}{\left(y_i-\theta_i\right)^2},
\end{equation}
where $N_\theta$ denotes the number of parameters (3 or 7), $\theta_i$ the value of the $i^\text{th}$ parameter with the following normalisation: (a) $\Om$ retain its original value, (b) $\s8$ is replaced by $S_8$, (c) $A_\mathrm{IA}$ is divided by a factor of $10$, and (d) the baryonic parameters ($M_\mathrm{c}$, $M_{1,0}$, $\eta$, $\beta$) are scaled such that their prior ranges map to $[0.0,0.5]$ after taking their logarithm. With normalisation, the value ranges of all parameters are around 0.5, which helps the convergence of the network.

\subsubsection{Training}
\label{sec:cnn-training}

For the analyses without baryonic effects or with a fixed baryonic model ($N_\theta=3$), we generate 800 ray-tracing realisations per cosmology ($1.2\ee6$ maps in total) for training. For the analyses with free baryonic parameters, we generate 1600 ray-tracing realisations  ($2.4\ee6$ maps in total), where the baryonic parameters are randomly chosen according to a Sobol sequence \citep{sobol1967} in the 4-dimensional hypercube such that each parameter follows a log-uniform distribution across its prior range. The post processing steps, including the addition of shape noise, IA, $\Delta m$, and PSF error, are applied on the fly, meaning that different maps will be generated during each visit to the same ray-tracing realisation.

The network is trained with the Adam optimiser \citep{kingma2014} with $\beta_1=0.9$ and $\beta_2=0.999$ for eight epochs: the first five epochs use a learning rate of $\alpha=10^{-3}$, then the learning rate is decreased by a factor 10 at the beginning of each of the rest three epochs. We have verified that for the networks used in this study, these numbers of training epochs are sufficient for the training losses to converge at each learning rate stage. In each training step, the network takes batches of 256 tomographic convergence maps distributed across four graphics processing units (GPUs). The training runs were performed on the TACC Frontera cluster using four Quadro RTX 5000 GPUs, with each run taking three to five hours depending on network hyper-parameters.

\subsection{Cross power spectra and tomographic peak counts}
\label{sec:cross-power-spectra}

In addition to the CNN, we also perform analyses with cross power spectra and tomographic peak counts on the same set of convergence maps, so that we can fairly compare their qualities.

The cross power spectrum between the convergence maps of two redshift bins $j$ and $k$ ($1\le j\le k\le4$) is measured by:
\begin{equation}
C^{jk}(i)=\frac{1}{N_{\mathrm{mode},i}}\sum_{\ell_{i}\le|\boldsymbol{\ell}|<\ell_{i+1}}C^{jk}(\boldsymbol{\ell}),
\end{equation}
where $i=1,\cdots,8$ denote the indices of the multipole bins,  $\ell_{i}=10^{2.4+i/10}$ the boundaries of the bins, $N_{\mathrm{mode},i}$ the number of multipole modes in the $i^\mathrm{th}$ bin, $C^{jk}(\boldsymbol{\ell})$ the cross power spectrum at multipole $\boldsymbol{\ell}$. The multipole range of the measured cross power spectra is $316<\ell<1995$, comparable to that in \citet{hikage2019}. Across all ten pairs of the redshift bins, the cross power spectra of a tomographic convergence map produces a data vector with 80 numbers.

The peak counts of a convergence map in a redshift bin is defined to be the histogram of all peak values in the map, where a pixel is a peak if and only if its $\kappa$ value is greater than all eight of its neighbours. We smooth the maps by a Gaussian kernel with radius $\sigma_\mathrm{G}=2\unit{arcmin}$ to remove small scale fluctuations. By inspecting the convergence maps at the fiducial cosmology, we find the typical standard deviations of $\kappa$ across the maps are $\sigma_\kappa=0.015,0.016,0.019,0.025$ in the four redshift bins (low to high) respectively. Then number of peaks are counted in 20 equally-spaced $\kappa$ bins in the range of $-1\,\sigma_\kappa^{(b)}\le\kappa\le 4\,\sigma_\kappa^{(b)}$, with the width of each bin being $0.25\,\sigma_\kappa^{(b)}$ for the $b^\text{th}$ redshift bin.

\section{Parameter inference}
\label{sec:parameter-inference}

In this section, we will describe in detail the process of inferring the cosmological parameters and, in some analyses, the baryonic parameters using summary statistics. We infer the parameters by comparing the target data vector, which can come from the simulations or the real data, to the model from our simulation pipeline. Note that this procedure is different from \citet{hikage2019} and \citet{hamana2020}, where they primarily rely on theoretical predictions to construct the model.

We first generate 800 ray-tracing realisations per cosmology from our pipeline, for each variation of baryonic modelling (no baryon, fiducial parameters, and free parameters). We note that these set of realisations are different from those being used to train the CNN in Section~\ref{sec:cnn-training}. Then, we apply a summary statistic to all of the simulated tomographic convergence maps and compute one data vector per realisation:
\begin{equation}
\boldsymbol{y}_{r}(\boldsymbol\theta)=\frac{1}{N_\mathrm{field}}\sum_{s=1}^{N_\mathrm{field}}f\left(\kappa_{r}^{(s)}(\boldsymbol\theta,\boldsymbol{x})\right),
\end{equation}
where $N_\mathrm{field}=19$ denotes the number of sub-fields, $f$ the summary statistic, and $\kappa_{r}^{(s)}(\boldsymbol\theta,\boldsymbol{x})$ the tomographic convergence map for the $r^\text{th}$ realisation and $s^\text{th}$ sub-field in the model with parameters $\boldsymbol\theta$.
We evaluate $\boldsymbol{y}_r(\boldsymbol{\theta})$ at following discrete $\theta$ values:
\begin{itemize}
\item for all $r$, $(\Om,\s8)$ is one of the simulated cosmologies in Section~\ref{sec:nbody};
\item for all $r$, $A_\mathrm{IA}=-3.0, -2.5, -2.0, \cdots, 3.0$;
\item for a specific $r$, the four baryonic parameters take the values of the $r^\text{th}$ term in the Sobol sequence as in Section~\ref{sec:cnn-training}.
\end{itemize}
Additionally, the target data vector $\boldsymbol{y}$ is calculated in the same way.

We model the likelihood of observing the data vector $\boldsymbol{y}$ given the parameters $\boldsymbol{\theta}$ as a multi-dimensional normal distribution:
\begin{gather}
p(\boldsymbol{y}|\boldsymbol{\theta})\propto\frac{1}{\sqrt{\det\boldsymbol{\mathsf{C}}(\boldsymbol\theta)}}\exp\left(-\frac{1}{2}\Delta\boldsymbol{y}^\mathrm{T}\tilde{\boldsymbol{\mathsf{C}}}^{-1}\Delta\boldsymbol{y}\right),  \label{eqn:likelihood}\\
\Delta\boldsymbol{y}=\boldsymbol{y}-\langle\boldsymbol{y}_r\rangle(\boldsymbol{\theta}),  \label{eqn:delta-y}\\
\tilde{\boldsymbol{\mathsf{C}}}^{-1}\equiv\frac{N_\mathrm{sim}-N_\mathrm{theta}-2}{N_\mathrm{sim}-1}\boldsymbol{\mathsf{C}}(\boldsymbol\theta)^{-1}, \label{eqn:covariance}
\end{gather}
where $\boldsymbol{\mathsf{C}}(\boldsymbol\theta)$ is the covariance of $\boldsymbol{y}_{r}(\boldsymbol\theta)$, $N_\mathrm{sim}=800$ the number of simulated realisations, and $N_\mathrm{theta}$ the number of dimensions in the data vector (3 or 7 for the CNN and 80 for power spectra and peak counts).  The summary statistics of 800 realisations from a single $500\hMpc$ $N$-body simulation can be considered statistically independent \citep{petri2016}, and the pre-factor in Equation~\eqref{eqn:covariance} is the Anderson-Hartlap correction factor \citep{hartlap2007} that makes the precision matrix $\tilde{\boldsymbol{\mathsf{C}}}^{-1}$ unbiased. Previous studies \citep[see e.g.][]{lin2015,gupta2018,lu2022} have shown that the data vectors following multi-dimensional normal distributions is a reasonable assumption for parameter inference from either the peaks counts or from the outputs of the CNNs.

Since $\boldsymbol{y}_r(\boldsymbol{\theta})$ are only available at discrete $\boldsymbol{\theta}$ values, we need to interpolate between them so that $\langle\boldsymbol{y}_r\rangle$ and $\boldsymbol{\mathsf{C}}$ can cover the entire parameter space. To avoid numerical instability, we interpolate the elements in the Cholesky decomposition of the covariance matrix, and we can reconstruct the interpolated covariance matrix by
\begin{equation}
\hat{\boldsymbol{\mathsf{C}}}(\boldsymbol\theta)=\hat{\boldsymbol{\mathsf{L}}}(\boldsymbol\theta)\hat{\boldsymbol{\mathsf{L}}}^\mathrm{T}(\boldsymbol\theta),
\end{equation}
where $\hat{\boldsymbol{\mathsf{L}}}(\boldsymbol\theta)$ is the interpolated Cholesky decomposition. For the model without free baryonic parameters, we find that a second-order polynomial fit to each element with respect to $A_\mathrm{IA}$ at each cosmology is sufficient:
\begin{equation}
v_i(A_\mathrm{IA})=a_i+b_i A_\mathrm{IA}+c_i A_\mathrm{IA}^2+\epsilon,
\end{equation}
where $v_i$ is one of the interpolated values, $(a_i,b_i,c_i)$ the polynomial coefficients, and $\epsilon$ a small unspecified error term. Then, $a_i$, $b_i$, and $c_i$ are interpolated linearly between the cosmologies with Delaunay triangulation (see Figure~\ref{fig:cosmologies}). For the model with free baryonic parameters, additional linear terms with respect to the baryonic parameters are added to the polynomial to fit $\langle\boldsymbol{y}_r\rangle$
\begin{eqnarray}
\nonumber v_i(A_\mathrm{IA},M_\mathrm{c},M_{1,0},\eta,\beta)=a_i+b_i A_\mathrm{IA}+c_i A_\mathrm{IA}^2 \\
+d_i M_\mathrm{c}+e_i M_{1,0}+f_i \eta+g_i \beta+\epsilon,
\end{eqnarray}
where the polynomial coefficients are interpolated between the cosmologies. For the covariance matrices, we reuse the values interpolated from the fiducial baryonic model, which means that the covariance is constant with respect to the baryonic parameters.

With the calculation of the likelihood function $p(\boldsymbol{y}|\boldsymbol{\theta})$ ready, we derive the posterior distribution with Bayes' theorem
\begin{equation}
p(\boldsymbol{\theta}|\boldsymbol{y})\propto p(\boldsymbol{y}|\boldsymbol{\theta})\,p(\boldsymbol{\theta}),
\end{equation}
where $p(\boldsymbol{\theta})$ is the prior distribution, as summarised in Table~\ref{tab:prior}. We sample the posterior distribution with a Monte Carlo Markov (MCMC) chain of $5\ee{6}$ steps, which is sufficient for the chains to converge.

\section{Cosmological constraints}
\label{sec:constraints}

\begin{table*}[!t]
\centering
\begin{tabular}{llllllll}
\hline
Name & Target data vector & Summary    & Baryonic & IA & $\Delta z$,$\Delta m$, & Other      & Section \\
     & calculated with & Statistics & model    &    & PSF                    & treatments & \\ \hline
\texttt{mock}            & T17 mock catalogues  & All & None     &            &            & & \ref{sec:constraints-mock}      \\
\texttt{no-systematics} & Our fiducial catalogues & All & None     &           &    & & \ref{sec:constraints-mock}       \\
\texttt{hsc-default}     & HSC catalogue    & All & None     & \checkmark & \checkmark & & \ref{sec:constraints-hsc}      \\
\texttt{hsc-no-ia}       & HSC catalogue    & All & None     &            & \checkmark & & \ref{sec:constraints-hsc}      \\
\texttt{hsc-fid-baryon}  & HSC catalogue    & All & Fiducial & \checkmark & \checkmark & & \ref{sec:constraints-hsc}      \\
\texttt{hsc-free-baryon} & HSC catalogue    & All & Free     & \checkmark & \checkmark & & \ref{sec:constraints-hsc}      \\
\texttt{bootstrap}       & Our fiducial catalogues & All & None     & \checkmark & \checkmark & & \ref{sec:discussion}  \\
\texttt{gaussian}        & Our fiducial catalogues & CNN & None     & \checkmark & \checkmark & Convert to GRFs & \ref{sec:non-gaussian}  \\
\texttt{smoothing}      & Our fiducial catalogues & CNN & None     & \checkmark & \checkmark & Various smoothing scales & \ref{sec:smoothing} \\ \hline
\end{tabular}
\caption{The configurations of the MCMC sampling runs.}
\label{tab:configurations}
\end{table*}

In this section, we present the cosmological and baryonic constraints using our pipeline in Section~\ref{sec:simulation} and the parameter inference method in Section~\ref{sec:parameter-inference}. In addition to the real HSC catalogue, we will also use T17 mock catalogues (see Section~\ref{sec:mock-catalogues}) and our own simulated catalogues from the fiducial cosmology to construct the target data vector ($\boldsymbol{y}$ in Equation~\eqref{eqn:delta-y}), so that we can validate our model and testing methods and effects. Hereafter, we will use the terms ``with the HSC catalog'', ``with T17 mock catalogues'', and ``with our fiducial catalogues'' to represent those target data vector choices respectively. The specification of all MCMC sampling runs are summarised in Table~\ref{tab:configurations}. In this work, we characterise the constraint of a parameter by the $16^\text{th}$ percentile $a$, median $b$, and $84^\text{th}$ percentile $c$ of the posterior distribution in the form of $b_{a-b}^{c-b}$, and the statistical error refers to the difference $c-a$, which corresponds to the difference between $-1\sigma$ and $+1\sigma$ in a normal distribution.

\subsection{Constraints with T17 mock catalogues}
\label{sec:constraints-mock}

\begin{figure}[!t]
\centering
\includegraphics[width=7.5cm]{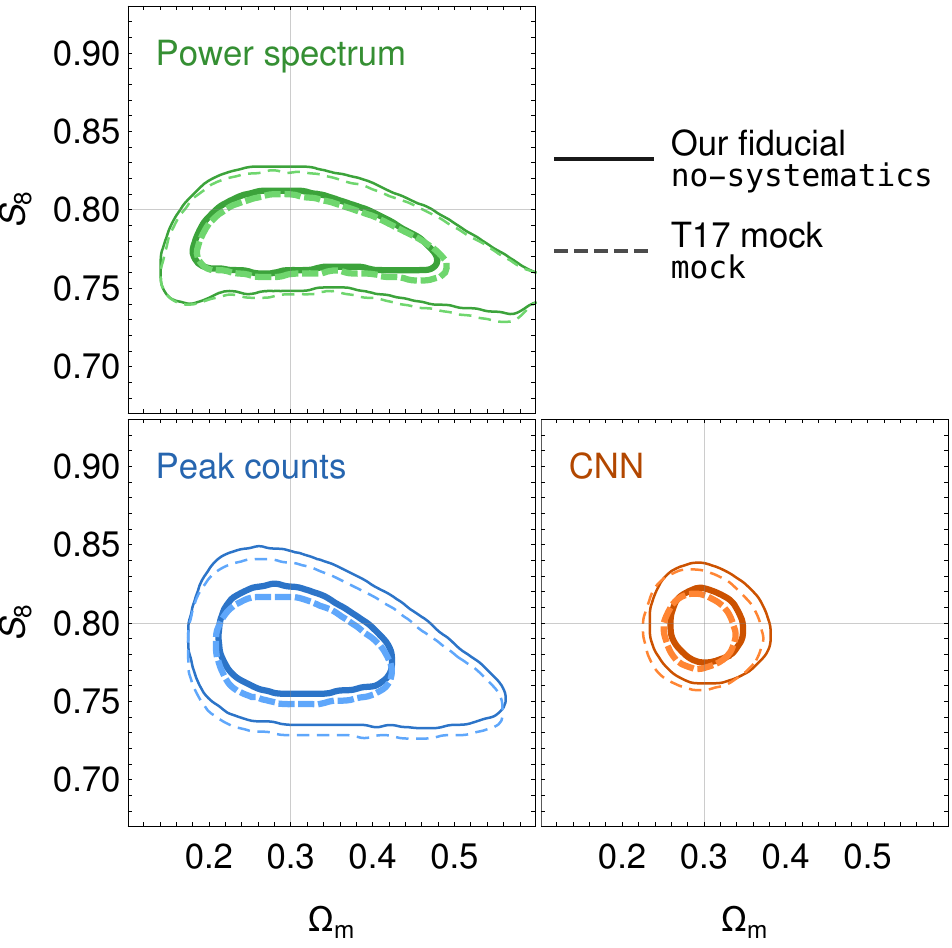}
\caption{Posterior distributions with the target data vectors calculated from our own catalogues at the fiducial cosmology ($\Om=0.3$, $S_8=0.8$) and those from T17 mock catalogues ($\Om=0.279$, $S_8=0.791$). The solid and dashed lines shows the 68 percent and 95 percent credible contours.}
\label{fig:corner-mock}
\end{figure}

First, we compare our simulation pipeline to the T17 mock simulations by constraining the cosmological parameters with the T17 mock catalogues generated in Section~\ref{sec:mock-catalogues}. Specifically, we calculate the data vector for each mock catalogue realisation and use the average data vector as the target data vector in Equation~\eqref{eqn:delta-y}.\footnote{A more rigorous comparison can be made by setting the target data vector to the data vector from each realisation and summarising the properties of the posterior distributions, instead of using a single averaged data vector.} We show the posterior distributions in Figure~\ref{fig:corner-mock}, together with the constraints where the data vector is calculated with our own simulated catalogues at the fiducial cosmology. Here, we use the configurations \texttt{mock} and \texttt{no-systematics}, where the IAs, baryons, and other systematics are not included, because non of these are added to T17 mock catalogues. To make the contours easier to read, we plot $S_8$ instead of $\s8$ in Figure~\ref{fig:corner-mock} and hereafter.

We find that the posterior distribution from our fiducial catalogues and that from T17 simulations are very similar other than small translations. In terms of the marginal distributions of $S_8$, the median $S_8$ values from the T17 mock catalogues are lower than that from our fiducial cosmology by $3.5\ee{-3}$, $6.4\ee{-3}$, and $3.7\ee{-3}$ for the power spectrum, peak counts, and CNN respectively. In comparison, the actual $S_8$ difference from cosmological parameters of the simulations is $\Delta S_8=9.2\ee{-3}$---the discrepancy is very small relative to the statistical error of $S_8$. We conclude that our model prediction and parameter inference pipelines passed this validation step.

Additionally, we note that the contours from our fiducial catalogues (the solid contours in Figure~\ref{fig:corner-mock}) are in general not centred around $\Om=0.3$ and $S_8=0.8$. In particular for the power spectrum, the maximum likelihood is reached at $(\Om=0.290, S_8=0.786)$. The reason is that the determinant of the covariance matrix varies across the cosmological parameter space, which affects the posterior probability in the prefactor of Equation~\eqref{eqn:likelihood}. This phenomenon is especially significant for the power spectrum because the determinant decreases by a factor of two when $S_8$ increases from 0.796 to 0.804 (a one percent increase). 

\subsection{Constraints with the HSC catalogue}
\label{sec:constraints-hsc}

\begin{figure}[!t]
\centering
\includegraphics[width=8.3cm]{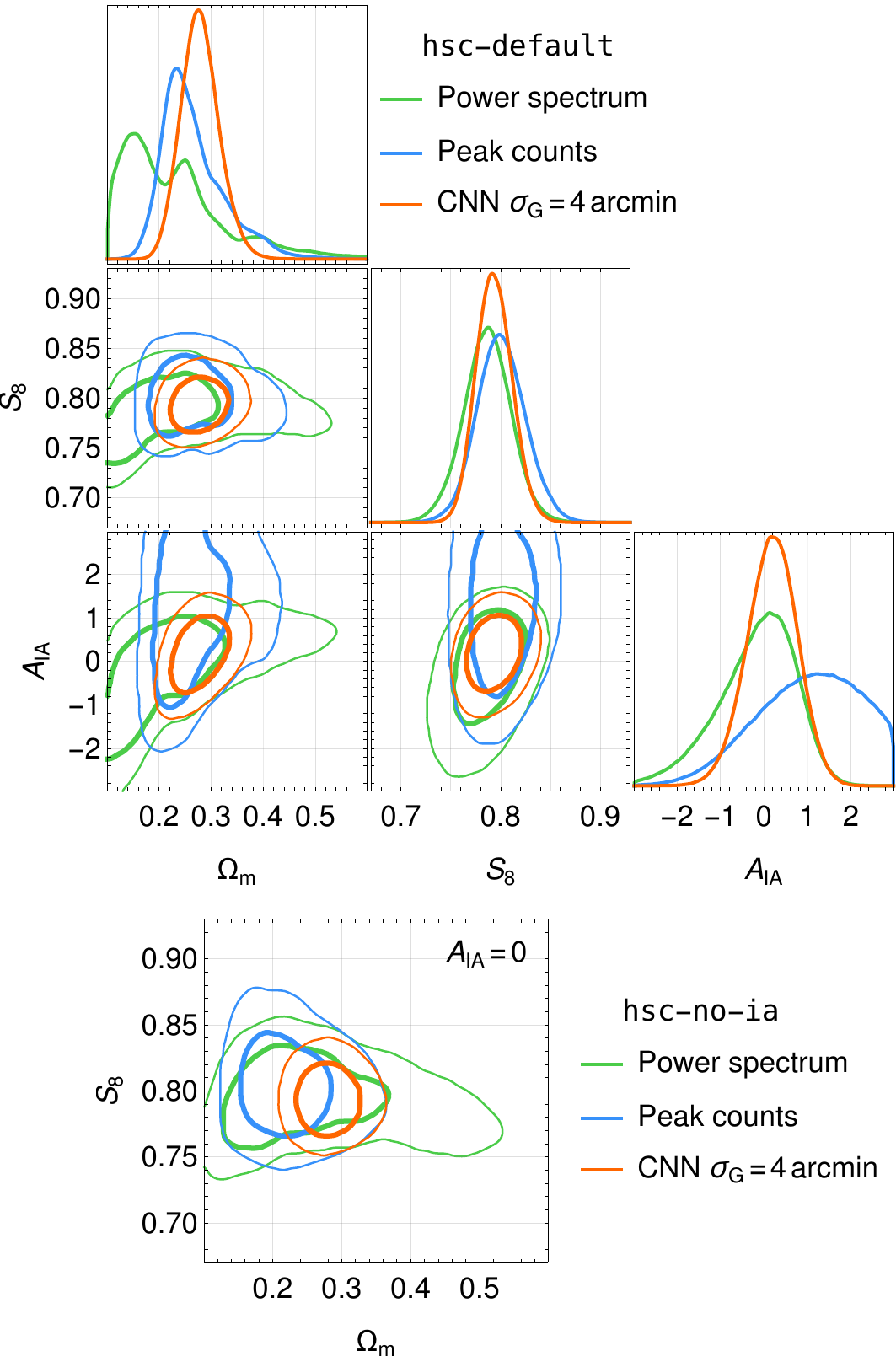}
\caption{Posterior distributions based on the HSC data in the model without baryons (top panel), and its conditional distribution with $A_\mathrm{IA}=0$ (bottom panel).}
\label{fig:corner-hsc-default}
\end{figure}

\begin{figure}[!t]
\centering
\includegraphics[width=8.0cm]{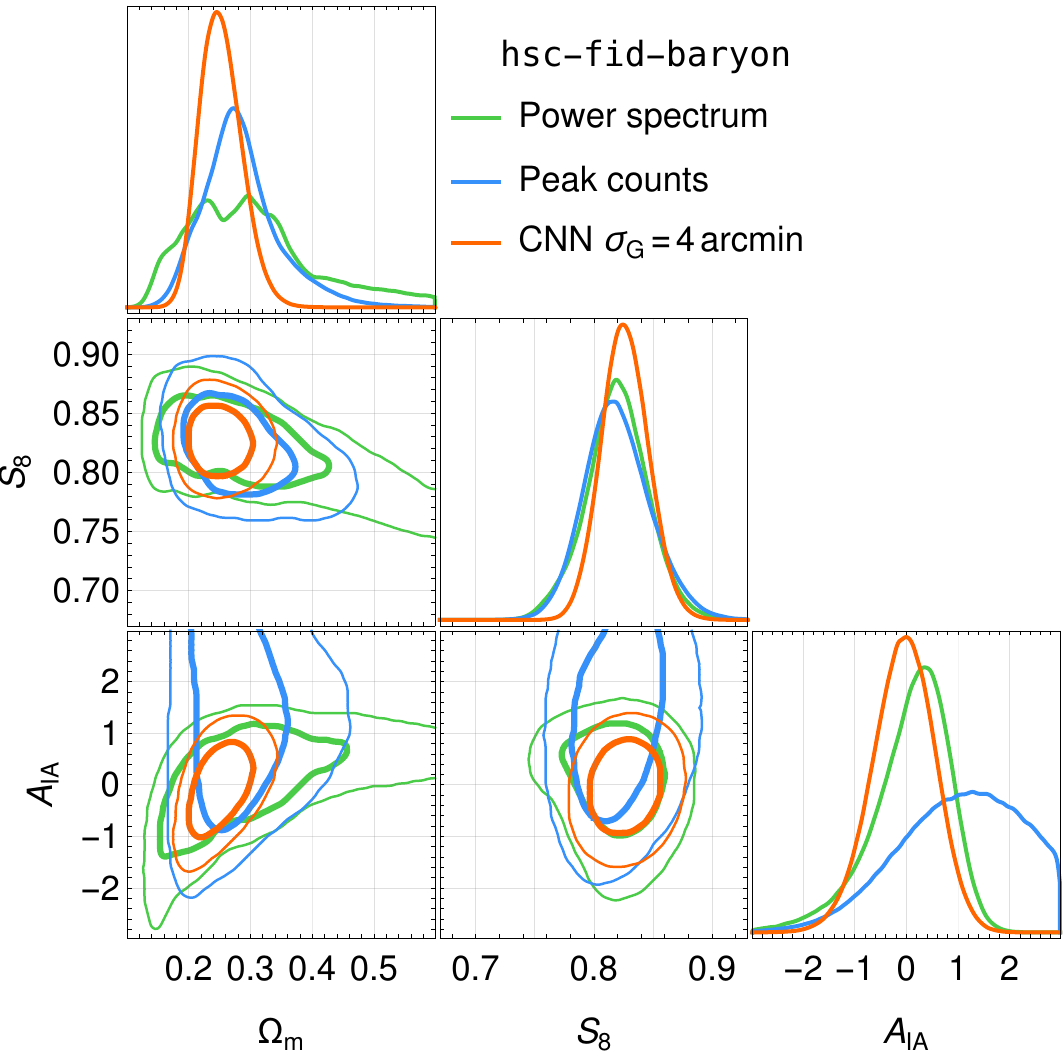}
\caption{Posterior distributions based on the HSC data in the model with fiducial baryonic parameters.}
\label{fig:corner-hsc-fid-baryon}
\end{figure}

\begin{figure*}[!t]
\centering
\includegraphics[width=17.0cm]{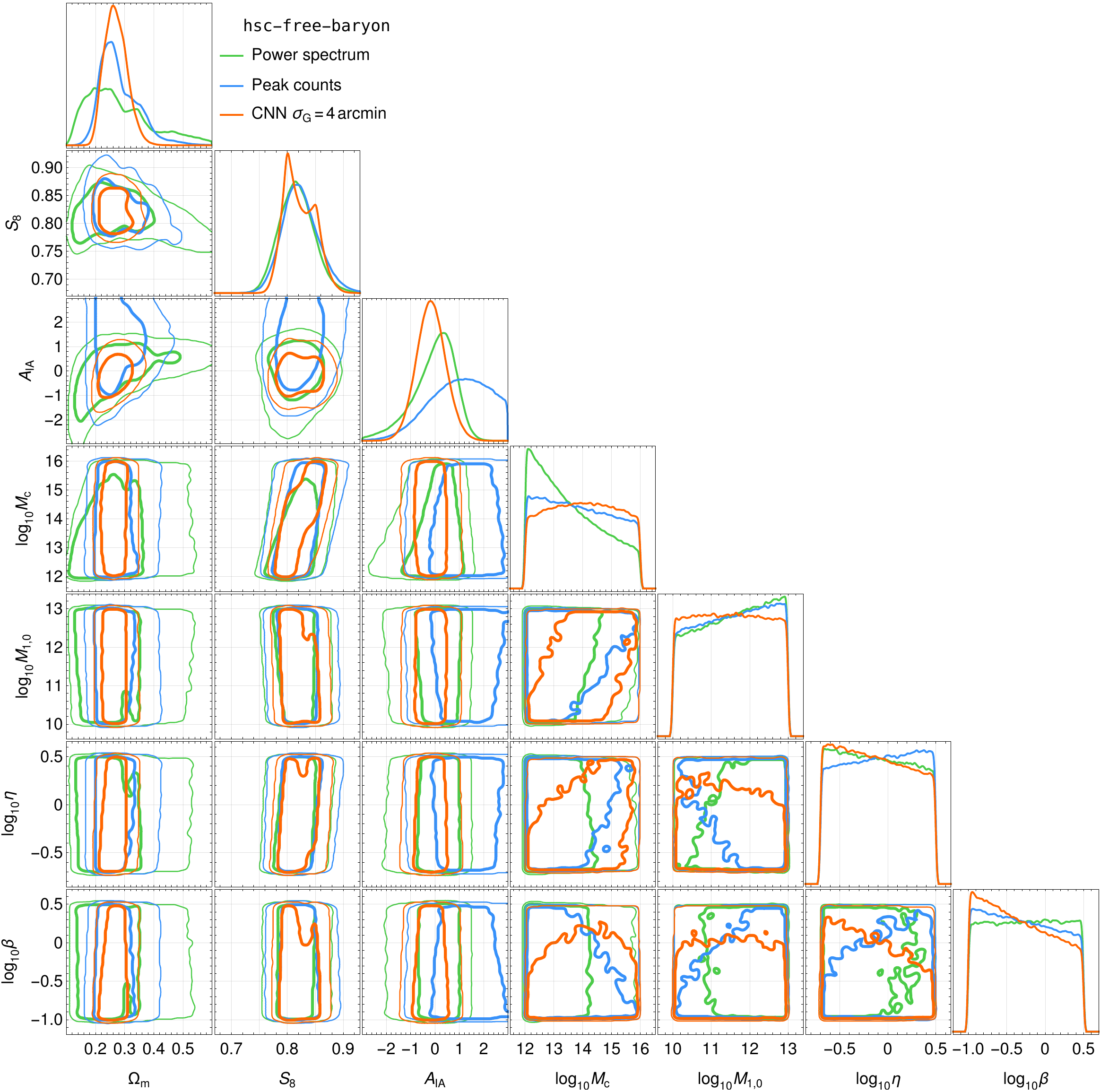}
\caption{Posterior distributions based on the HSC data in the model with free baryonic parameters.}
\label{fig:corner-hsc-free-baryon}
\end{figure*}

\begin{table*}[!t]
\centering
\begin{tabular}{llccc}
\hline
Model & Summary statistic & $\Om$ & $S_8$ & $A_\mathrm{IA}$ \\ \hline
\multirow{3}{*}{Default (no baryon)}
& Power spectrum & ${0.211}_{-0.072}^{+0.105}\;(1.00)$ & ${0.787}_{-0.023}^{+0.023}\;(1.00)$ & ${-0.07}_{-0.99}^{+0.75}\;(1.00)$ \\
& Peak counts    & ${0.254}_{-0.042}^{+0.072}\;(0.64)$ & ${0.800}_{-0.023}^{+0.025}\;(1.03)$ & ${1.10}_{-1.31}^{+1.15}\;(1.41)$ \\
& CNN            & ${0.278}_{-0.035}^{+0.037}\;(0.41)$ & ${0.793}_{-0.017}^{+0.018}\;(0.76)$ & ${0.20}_{-0.58}^{+0.55}\;(0.65)$ \\ \hline
\multirow{3}{*}{No IA ($A_\mathrm{IA}=0$)}
& Power spectrum & ${0.235}_{-0.063}^{+0.111}\;(1.00)$ & ${0.794}_{-0.022}^{+0.022}\;(1.00)$ & -- \\
& Peak counts    & ${0.218}_{-0.039}^{+0.049}\;(0.51)$ & ${0.803}_{-0.024}^{+0.025}\;(1.12)$ & -- \\
& CNN            & ${0.281}_{-0.029}^{+0.031}\;(0.35)$ & ${0.793}_{-0.017}^{+0.018}\;(0.80)$ & -- \\ \hline
\multirow{3}{*}{Fiducial baryonic model}
& Power spectrum & ${0.290}_{-0.087}^{+0.115}\;(1.00)$ & ${0.820}_{-0.026}^{+0.025}\;(1.00)$ & ${0.18}_{-0.82}^{+0.61}\;(1.00)$ \\
& Peak counts    & ${0.278}_{-0.051}^{+0.063}\;(0.57)$ & ${0.819}_{-0.026}^{+0.029}\;(1.08)$ & ${1.11}_{-1.28}^{+1.14}\;(1.70)$ \\
& CNN            & ${0.250}_{-0.031}^{+0.036}\;(0.33)$ & ${0.826}_{-0.019}^{+0.020}\;(0.77)$ & ${-0.04}_{-0.61}^{+0.58}\;(0.83)$ \\ \hline
\multirow{3}{*}{Free baryonic model}
& Power spectrum & ${0.259}_{-0.083}^{+0.139}\;(1.00)$ & ${0.816}_{-0.030}^{+0.031}\;(1.00)$ & ${0.08}_{-0.98}^{+0.68}\;(1.00)$ \\
& Peak counts    & ${0.268}_{-0.047}^{+0.082}\;(0.58)$ & ${0.822}_{-0.029}^{+0.034}\;(1.04)$ & ${1.07}_{-1.32}^{+1.19}\;(1.51)$ \\
& CNN            & ${0.268}_{-0.036}^{+0.040}\;(0.35)$ & ${0.819}_{-0.024}^{+0.034}\;(0.95)$ & ${-0.16}_{-0.58}^{+0.59}\;(0.71)$ \\ \hline

\end{tabular}
\caption{Cosmological constraints with the HSC catalogue in various models. The numbers in the parentheses show the statistical errors of the constraints relative to those by the power spectrum.}

\label{tab:constraints}
\end{table*}

In our fiducial analysis (\texttt{hsc-default}), we include the all of systematic effects except the baryonic effects, as in \citet{hikage2019}. The posterior distributions of $\Om$, $\s8$, and $A_\mathrm{IA}$ for the three methods are shown in Figure~\ref{fig:corner-hsc-default}. Note that we again replace $\s8$ by $S_8$ in the plots for the clarity of the contours. We find $S_8=0.787_{-0.023}^{+0.023}$ from the power spectrum, $S_8=0.800_{-0.023}^{+0.025}$ from the peak counts, and $S_8=0.793_{-0.018}^{+0.017}$ from the CNN. These values fall between the results in previous HSC studies---$S_8(\alpha=0.5)=0.780_{-0.033}^{+0.030}$ \citep{hikage2019} and $S_8=0.823_{-0.028}^{+0.032}$ \citep{hamana2020}. But we find that the statistical uncertainty for the power spectrum method in this work is 20--30 percent smaller than that by \citet{hikage2019}. We suspect that this is because \citet{hikage2019} have modelled eight parameters compared to only three in our fiducial analysis. Although the additional parameters does not appear to be correlated with $S_8$, nor are they constrained better than their priors, they can contribute to the statistical error of $S_8$.

While the posterior distributions of the three methods are in very good agreement, the CNN achieves the smallest statistical error for each of the parameters. Specifically, the CNN gives $S_8=0.793_{-0.018}^{+0.017}$, $\Om=0.278_{-0.035}^{+0.037}$ and $A_\mathrm{IA}=0.20_{-0.58}^{+0.55}$, which are better than the power spectrum by a factor of 1.3, 2.5, and 1.5 respectively. We notice a slight multimodality in the marginal distribution of $\Om$ for the power spectrum. We speculate that this is likely caused by the cosmology grid (in Figure~\ref{fig:cosmologies}) being relatively coarse along the $\Om$ direction, resulting in small inaccuracies from interpolation. This can be verified, and improved upon, by increasing the number of simulations in future works.

Figure~\ref{fig:corner-hsc-default} also shows the conditional distributions with $A_\mathrm{IA}=0$ (\texttt{hsc-no-ia}). Compared to the fiducial analysis, the marginal distributions of $S_8$ almost remain the same---the median values only shift by less than $\sim0.3\,\sigma$, but the constraints of $\Om$ have changed slightly due to the degeneracy between $\Om$ and $A_\mathrm{IA}$.

Figure~\ref{fig:corner-hsc-fid-baryon} shows the posterior distributions in the fiducial baryonic model (\texttt{hsc-fid-baryon}). We find that the inclusion of baryons shifts the constraints on $S_8$ higher by 0.02--0.03. Such differences are expected since the overall effect of baryons is to lower the matter fluctuation on small physical scales, which is negatively correlated with a higher value of $S_8$. 

Figure~\ref{fig:corner-hsc-free-baryon} shows the posterior distributions in the model with free baryonic parameters (\texttt{hsc-free-baryon}), which has seven free parameters in total. The cosmological constraints by the CNN are $\Om=0.268_{-0.036}^{+0.040}$, $S_8=0.819_{-0.024}^{+0.034}$, and $A_\mathrm{IA}=-0.16_{-0.58}^{+0.59}$. Other than a slight degeneracy between $S_8$ and $M_\mathrm{c}$ and the power spectrum preferring the model with low $M_\mathrm{c}$, none of the methods can extract enough information from the maps to constrain the baryonic parameters. \citet{lu2022} have shown that these methods should marginally constrain baryonic parameters only with the full HSC survey area according to their simulation-based forecast, so it is not surprising that we can not constrain baryonic parameters with the smaller first-year survey area. With the impact of free baryonic parameters, the statistical errors of $S_8$ increase by 15--30 percent for the three methods relative to those in fiducial baryonic model, but the median $S_8$ values are almost unchanged.

The credible intervals for the above HSC constraints are summarised in Table~\ref{tab:constraints}.

\section{Discussion}
\label{sec:discussion}

\subsection{CNN hyper-parameters}
\label{sec:hyperparameters}

\begin{table}[!t]
\centering
\begin{tabular}{cccccc}
\hline
$n_\mathrm{block}$ & $n_\mathrm{ch}$ & $\sigma_\mathrm{G}$ & \multicolumn{3}{c}{Statistical error} \\
 & & & $\Om$ & $S_8$ & $A_\mathrm{IA}$ \\ \hline
 5 & 32 & $4'$ & 0.0778 & 0.0351 & 1.15 \\
10 & 32 & $4'$ & 0.0769 & 0.0351 & 1.15 \\
15 & 32 & $4'$ & 0.0782 & 0.0354 & 1.15 \\
 5 & 64 & $4'$ & 0.0776 & 0.0354 & 1.14 \\
\textbf{10} & \textbf{64} & $\mathbf{4'}$ & \textbf{0.0767} & \textbf{0.0355} & \textbf{1.16} \\
15 & 64 & $4'$ & 0.0756 & 0.0351 & 1.16 \\
 5 & 96 & $4'$ & 0.0779 & 0.0355 & 1.15 \\
10 & 96 & $4'$ & 0.0761 & 0.0349 & 1.15 \\
15 & 96 & $4'$ & 0.0765 & 0.0353 & 1.13 \\ \hline
10 & 64 & $8'$ & 0.0918 & 0.0389 & 1.20 \\
\textbf{10} & \textbf{64} & $\mathbf{4'}$ & \textbf{0.0767} & \textbf{0.0355} & \textbf{1.16} \\
10 & 64 & $2'$ & 0.0696 & 0.0338 & 1.13 \\
10 & 64 & $1'$ & 0.0696 & 0.0338 & 1.13 \\ \hline
\end{tabular}
\caption{The statistical errors of the parameters using different CNN architectures and smoothing scales. }
\label{tab:hyper-parameters}
\end{table}

In our fiducial analysis, we use the CNN architecture with $n_\mathrm{block}=10$ and $n_\mathrm{ch}=64$, and in this section, we train eight additional CNNs with combinations of $n_\mathrm{block}=5,10,15$ and $n_\mathrm{ch}=32,64,96$. In Table~\ref{tab:hyper-parameters}, we show the statistical error on $\Om$, $S_8$, and $A_\mathrm{IA}$ for all CNNs with our fiducial catalogues (\texttt{bootstrap}). We find that all CNN architectures we tested performs similarly, which means that those architectures are all complex enough to extract the information from the maps.

\subsection{Map smoothing scale}
\label{sec:smoothing}

\begin{figure*}[!t]
\centering
\includegraphics[width=16.0cm]{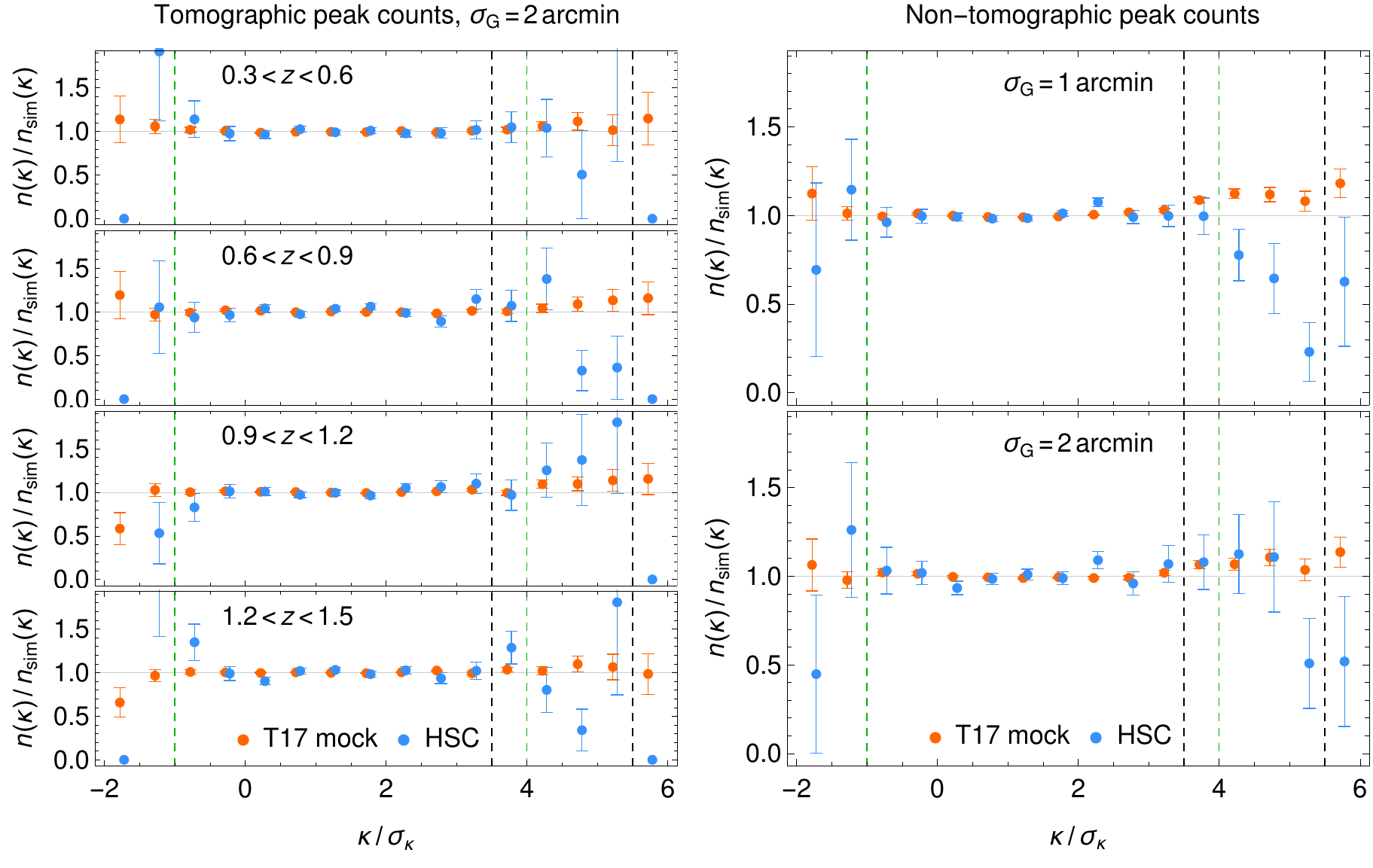}
\caption{Comparison between the HSC or T17 mock peak counts and our simulated peak counts. Left panels show the tomographic peak counts in four redshift bins, and the right panels show the non-tomographic peak counts with two smoothing scales. The error bars are calculated assuming Poisson distributions. The boundaries of the $\kappa/\sigma_\kappa$ ranges used in \citet{liu2022} and in our analyses are shown in black and green dashed lines respectively.}
%TODO: swap "HSC" and "T17 mock" in the plots
\label{fig:peak-counts}
\end{figure*}

In our fiducial analysis, the convergence maps taken by the CNN are smoothed by a Gaussian kernel with $\sigma_\mathrm{G}=4\unit{arcmin}$. Here, we train a few more CNNs on the convergence maps smoothed on other smoothing scales (\texttt{smoothing}): $\sigma_\mathrm{G}=1,2,\text{and }8\unit{arcmin}$. The statistical errors of the three parameters with those CNNs are in listed in Table~\ref{tab:hyper-parameters}. We find that decreasing the smoothing scale reduces statistical errors of the parameters, as the information on smaller scales becomes available. But when a certain smoothing scale is reached, the shape noise dominates the signal and decreasing the smoothing scale further will no longer improve the constraints. From our tests, this critical smoothing scale is $2\textrm{--}4\unit{arcmin}$ for CNNs: the reductions in statistical errors are small from $\sigma_\mathrm{G}=4\unit{arcmin}$ to $\sigma_\mathrm{G}=2\unit{arcmin}$ (9 percent for $\Om$ and 5 percent for $S_8$), and they are negligible from $\sigma_\mathrm{G}=2\unit{arcmin}$ to $\sigma_\mathrm{G}=1\unit{arcmin}$. A similar result is also found by \citet{hikage2019}, where extending from $\ell_\mathrm{max}=1900$ to $\ell_\mathrm{max}=3500$ only yields a 10 percent reduction on the error of $S_8$.

Although a smaller smoothing scale gives us tighter constraints, it makes the results more vulnerable to the systematic effects that are not well-understood or not accounted for. Here, we present an example where we find a discrepancy between the number of high peaks in the HSC data and that in our simulation. In Figure~\ref{fig:peak-counts}, we show the tomographic peak counts with $\sigma_\mathrm{G}=2\unit{arcmin}$ (used in our analyses) and the non-tomographic peak counts with $\sigma_\mathrm{G}=1\unit{arcmin}$ and $2\unit{arcmin}$. The number of peaks in each bin is normalized by that from the simulation assuming the cosmology constrained in Section~\ref{sec:constraints} ($\Om=0.254,\s8=0.800,A_\mathrm{IA}=1.10$ for HSC; $\Om=0.313,\s8=0.785,A_\mathrm{IA}=0$ for T17 mock). The non-tomographic maps are calculated on a higher resolution grid of $208\times208$ due to a higher galaxy number density on each map. We find that with $\sigma_\mathrm{G}=1\unit{arcmin}$ and $\kappa/\sigma_\kappa>4$, the non-tomographic peak counts from the HSC data are consistently lower than those from our simulations. This effect for $\sigma_\mathrm{G}=2\unit{arcmin}$, if there is any, has a much lower statistical significance. We speculate that this behaviour is caused by us ignoring the boost factor and dilution effect due to the clustering of the galaxies, which are two of main systematics for high peaks. According to \citet{liu2022}, taking the boost factor and dilution effect into account can decrease the number of high peaks ($\kappa/\sigma_\kappa>4.5$) by about 30--50 percent in non-tomographic maps with $\sigma_\mathrm{G}=1.5\unit{arcmin}$, which is comparable to our results in Figure~\ref{fig:peak-counts}. While $\kappa$ cuts can be made to prevent those systematics from affecting our peak counts analyses, it is much harder to modify a $\sigma_\mathrm{G}=1\unit{arcmin}$ map to prevent a CNN from learning high peak features.

Take both factors above into consideration, we find a similar conclusion to \citet{hikage2019} that a smoothing scale of $\sigma_\mathrm{G}=4\unit{arcmin}$ ($\ell\lesssim2000$) is a conservative choice to give reliable constraints.

\subsection{Non-Gaussian information}
\label{sec:non-gaussian}

\begin{figure}[!t]
\centering
\includegraphics[width=8.0cm]{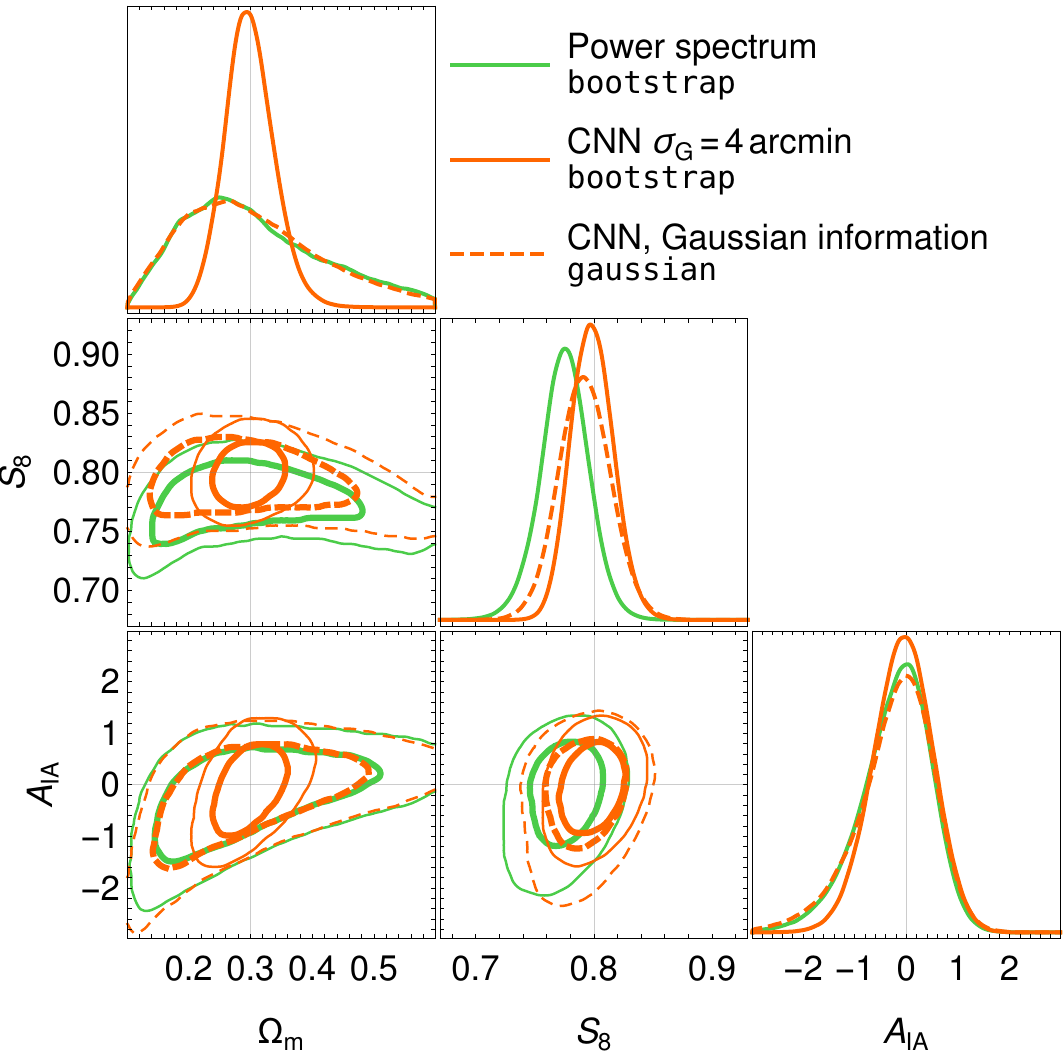}
\caption{Posterior distributions based on the fiducial cosmology. The dashed contours shows the constraints by the CNN on the maps with non-Gaussian information removed.}
\label{fig:corner-gaussian}
\end{figure}

The results in Section~\ref{sec:constraints} show that the CNN can produce tighter constraints than the power spectrum, especially for $\Om$, which suggests that the CNN can effectively extract non-Gaussian information from the maps. But since it is hard to fully analyse the internal working of the CNN, we need to perform tests to make sure that it does not underestimate the statistical errors through over-fitting to the training maps.

In Section~\ref{sec:parameter-inference}, we have mentioned that we can avoid over-fitting by using two separate sets of maps for training and for the construction of the likelihood function. Here, we perform a direct test to check the possibility of over-fitting using Gaussian random fields (GRFs), following a similar test by \citet{gupta2018}. In this test, we remove non-Gaussian information from each simulated tomographic convergence map by randomising the phase in the Fourier space; formally,
\begin{equation}
\hat{\kappa}^{(b)}(\boldsymbol\ell)\rightarrow e^{i\varphi(\boldsymbol\ell)}\hat{\kappa}^{(b)}(\boldsymbol\ell),
\end{equation}
where $\hat{\kappa}^{(b)}$ is the convergence map of the $b^\text{th}$ redshift bin in the Fourier space, and $\varphi$ a random phase across all redshift bins following a uniform distribution $\mathcal{U}(-\pi,\pi)$ that satisfies $\varphi(-\boldsymbol\ell)=-\varphi(\boldsymbol\ell)$. With the assumption that the weak lensing field is isotropic, this operation preserves the power spectrum---furthermore, the generated GRFs have the exact same amount of information that can be extracted through the power spectrum. Therefore, no summary statistics, including the CNN, should perform better than the power spectrum on those maps.

Figure~\ref{fig:corner-gaussian} shows the posterior distribution from the CNN with the GRFs (\texttt{gaussian}) along with that from the power spectrum using our fiducial catalogues (\texttt{bootstrap}). Apart from a small shift along the $S_8$ direction due to the varying determinant, these two contours are almost identical. Figure~\ref{fig:corner-gaussian} also shows the posterior distribution by the CNN on the original maps (\texttt{bootstrap}, with non-Gaussian information), where the contours are fully enclosed in the contours by the CNN on GRFs. These results suggest that this CNN architecture is working properly in extracting information---it does not underestimate the statistical errors on GRFs, and it can extract additional non-Gaussian information when available.

\subsection{Comparison with Planck 2018 results}
\label{sec:baryons}

\begin{figure}[!t]
\centering
\includegraphics[width=8.5cm]{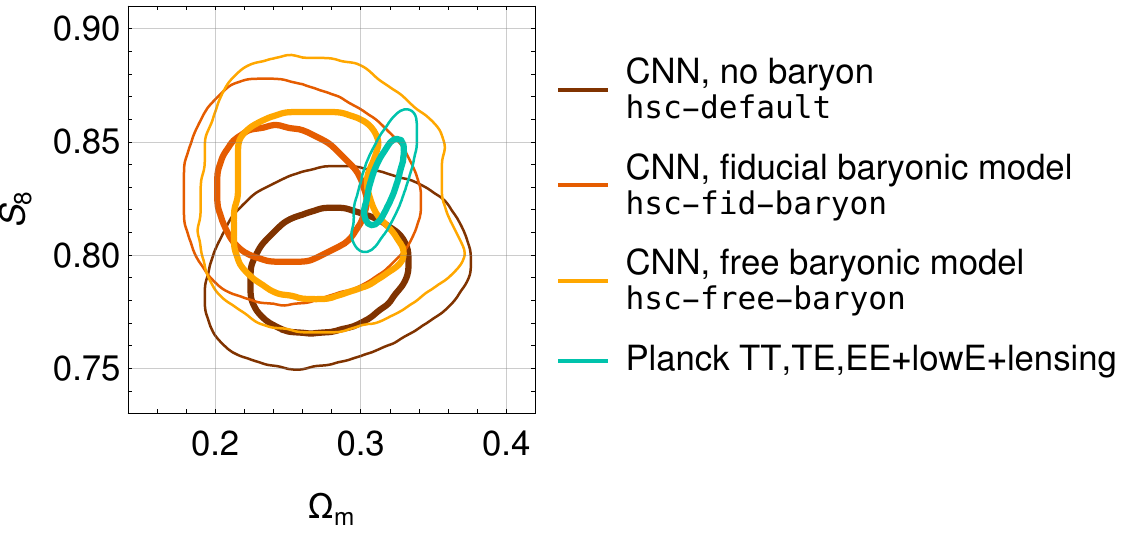}
\caption{The posterior distributions of $\Om$ and $S_8$ marginalised over other parameters, constrained by the CNNs with various baryonic models and by \citet{planck2018}.}
\label{fig:corner-planck}
\end{figure}

Many previous studies have found that the constraint on $S_8$ from WL is lower than that from the cosmic microwave background and have investigated the probable causes \citep{hikage2019,troxel2018,hildebrandt2017,joudaki2016,leauthaud2017}. Here, we show the constraints on $\Om$ and $S_8$ by the CNNs with various baryonic models and that by \citet{planck2018} (TT,TE,EE+lowE+lensing) in Figure~\ref{fig:corner-planck}. Contrary to the analysis in \citet{hikage2019}, where the baryonic effects are considered not important in constraining $S_8$, we find that they can be significant for the CNN (\texttt{hsc-fid-baryon} is higher by $\sim1.7\,\sigma$), because the $S_8$ shift measured in this work is higher than \citet{hikage2019} by $\sim50\text{ percent}$ while the statistical error on $S_8$ is 23 percent smaller. When compared to Planck 2018, we find that our constraint on $S_8$ without baryons is lower than Planck 2018 by $\sim2.2\,\sigma$, but with baryons, the discrepancy is much smaller ($\sim0.3\,\sigma$ for the fiducial baryonic model, $\sim0.5\,\sigma$ for the fiducial baryonic model). If additional cosmological parameters had been included in our study, the statistical error of $S_8$ could have been slightly larger (see Section~\ref{sec:constraints-hsc}), and the tension can be less pronounced in all cases. We also note that since the first-year HSC data can not provide constraints on the baryonic parameters, better prior distributions of the baryonic parameters may be obtained with the help of external observations, but this is beyond the scope of this study. Nevertheless, it is clear that baryonic effects account for at least a part, and possibly all of the discrepancy with Planck 2018 $S_8$ constraint. Similar results have been found by previous studies with other baryonic models: \citet{troxel2018} have shown that $S_8$ increases from $0.782_{-0.027}^{+0.027}$ to $0.798_{-0.028}^{+0.026}$ by including baryons, and \citet{hildebrandt2017} have shown that $S_8$ increases by $0.3\text{--}1.0\,\sigma$  by including baryons.

\section{Conclusions}
\label{sec:conclusions}

In this study, we have presented cosmological constraints from the first-year HSC data \citep{mandelbaum2018a} using CNNs as summary statistics, applied to lensing maps created from the public HSC shear catalogue. We have compared these to the constraints from the lensing convergence power spectrum and peak counts. The constraints come from our 4-bin tomographic analyses for the 8.4 million galaxies in the redshift range $0.3\le z\le 1.5$ and placed on 19 sub-fields with an area of $3\times3\unit{{deg}^2}$ each.

We have developed a simulation pipeline to produce mock convergence maps and create forward models with properties as close as possible to those of
the observed data. The pipeline starts with running 79 dark matter-only $N$-body simulations with varying $\Om$ and $\s8$. The simulated galaxy shear is ray-traced through the simulations, and we add shape noise to them based on the real catalogue. Lastly, we convert the real and simulated shears to convergence maps. During this process, intrinsic alignments, baryonic effects, photometric redshift estimation uncertainties, shear measurement biases, and PSF modelling errors are added.

We have trained the CNNs to predict the underlying cosmological, intrinsic alignment, and baryonic parameters from the simulated convergence maps. The training maps are smoothed by a Gaussian kernel with $\sigma_\mathrm{G}=4\unit{arcmin}$ so that the CNNs do not have access to the information on the smallest scales ($\ell\gtrsim2000$), which is consistent with the $\ell$ range of the power spectrum. We then used the trained networks to derive likelihood functions in the same way as for other summary statistics, and they are also applied to the HSC convergence maps to derive parameter estimates, which are treated as any other statistic. We have performed Bayesian inference with MCMC to sample the posterior.

In the $\Lambda$CDM model with two free cosmological parameters $\Om$ and $\s8$, we find $\Om=0.278_{-0.035}^{+0.037}$, $S_8=0.793_{-0.018}^{+0.017}$, and the IA amplitude $A_\mathrm{IA}=0.20_{-0.58}^{+0.55}$. By adding the baryonic effects with the fiducial parameters, we find $\Om=0.250_{-0.031}^{+0.036}$, $S_8=0.826_{-0.019}^{+0.020}$, and $A_\mathrm{IA}=-0.04_{-0.61}^{+0.58}$. In a baryonic model with free parameters, we find the baryonic parameters to be poorly constrained, and $\Om=0.268_{-0.036}^{+0.040}$, $S_8=0.819_{-0.024}^{+0.034}$, and $A_\mathrm{IA}=-0.16_{-0.58}^{+0.59}$ when marginalising over baryonic parameters. 

All statistical uncertainties of the constraints by the CNNs are smaller than those of the power spectrum and peak counts. Compared to the power spectrum, the statistical errors of $\Om$ by the CNNs is smaller by a factor of 2.5--3.0, and those of $S_8$ is smaller by 5--24 percent, showing the effectiveness of CNNs in extracting non-Gaussian information from the lensing signals. We note that if more cosmological parameters were included in our modelling, the statistical error of $S_8$ could have been larger, as suggested by a comparison between the power spectrum constraints in this study and by \citet{hikage2019}. We also find that the median value of $S_8$ is increased by 0.02--0.03 when we take baryons into consideration. With baryons, the $S_8$ discrepancy between our CNN constraints and Planck 2018 results is reduced from $\sim2.2\,\sigma$ to $0.3\text{--}0.5\,\sigma$.

\section*{Acknowledgements}

We thank Jia Liu and Jos\'e Zorrilla~Matilla for useful discussions and Cyrille Doux for comments that helped us clarify our manuscript. We acknowledge support by NASA ATP grant 80NSSC18K1093, the use of the NSF XSEDE facility Stampede2, for the simulations and data analysis in this study.

The Hyper Suprime-Cam (HSC) collaboration includes the astronomical communities of Japan and Taiwan, and Princeton University. The HSC instrumentation and software were developed by the National Astronomical Observatory of Japan (NAOJ), the Kavli Institute for the Physics and Mathematics of the Universe (Kavli IPMU), the University of Tokyo, the High Energy Accelerator Research Organization (KEK), the Academia Sinica Institute for Astronomy and Astrophysics in Taiwan (ASIAA), and Princeton University. Funding was contributed by the FIRST program from Japanese Cabinet Office, the Ministry of Education, Culture, Sports, Science and Technology (MEXT), the Japan Society for the Promotion of Science (JSPS), Japan Science and Technology Agency (JST), the Toray Science Foundation, NAOJ, Kavli IPMU, KEK, ASIAA, and Princeton University.

This paper makes use of software developed for the Large Synoptic Survey Telescope. We thank the LSST Project for making their code available as free software at  http://dm.lsst.org

The Pan-STARRS1 Surveys (PS1) have been made possible through contributions of the Institute for Astronomy, the University of Hawaii, the Pan-STARRS Project Office, the Max-Planck Society and its participating institutes, the Max Planck Institute for Astronomy, Heidelberg and the Max Planck Institute for Extraterrestrial Physics, Garching, The Johns Hopkins University, Durham University, the University of Edinburgh, Queen’s University Belfast, the Harvard-Smithsonian Center for Astrophysics, the Las Cumbres Observatory Global Telescope Network Incorporated, the National Central University of Taiwan, the Space Telescope Science Institute, the National Aeronautics and Space Administration under Grant No. NNX08AR22G issued through the Planetary Science Division of the NASA Science Mission Directorate, the National Science Foundation under Grant No. AST-1238877, the University of Maryland, and Eotvos Lorand University (ELTE) and the Los Alamos National Laboratory.

Based [in part] on data collected at the Subaru Telescope and retrieved from the HSC data archive system, which is operated by Subaru Telescope and Astronomy Data Center at National Astronomical Observatory of Japan.

\section*{Data availability}

The data underlying this article were accessed from Stampede2. The derived data generated in this research will be shared on reasonable request to the corresponding author.

\bibliography{main}{}
\bibliographystyle{aasjournal}

\end{document}